\documentclass[a4paper,11pt]{article}
\usepackage{jheppub} % for details on the use of the package, please
\usepackage[T1]{fontenc} % if needed

\usepackage{enumitem}

\title{\boldmath Broken Scaling Neutrino Mass Matrix and Leptogenesis based on A$_4$ Modular invariance}

\author[a,1]{Monal Kashav,\note{Corresponding author.}}
\author[a]{Surender Verma}

\affiliation[a]{Department of Physics and Astronomical Science\\
Central University of Himachal Pradesh\\
Dharamshala, India 176215}

\emailAdd{monalkashav@gmail.com}
\emailAdd{sverma@cuhimachal.ac.in}

\abstract{In this work, we have proposed a modular $A_4$ symmetric model of neutrino mass which, simultaneously,  explains observed baryon asymmetry of the Universe(BAU). In minimal extension of the standard model(SM) with two right-handed neutrinos we work in a supersymmetric framework. At Type-I seesaw level, the model predicts scaling in the neutrino mass matrix. In order to have correct low energy phenomenology, we propose two possible scenarios of scale-breaking in the neutrino mass matrix emanating from Type-I seesaw. Scenario-1 is based on the dimension-5 Weinberg operator whereas scenario-2 implements Type-II seesaw via scalar triplet Higgs superfields($\Delta,\bar{\Delta}$). Interestingly, the breaking patterns in both, otherwise dynamically different scenarios, are similar which can be attributed to the same charge assignments of superfields($\Delta,\bar{\Delta}$) and the Higgs superfield $H_u$ under modular $A_4$ symmetry. The breaking is found to be proportional to the Yukawa coupling of modular weight 10($Y_{1,1'}^{10}$). We, further, investigates the predictions of the model under scenario-2 (Type-I+II) for neutrino mass, mixings and matter-antimatter asymmetry of the Universe. The model predicts normal hierarchical neutrino masses and provide a robust range ($0.05-0.08$)eV for sum of neutrino masses($\sum m_{i}$). Lepton number violating $0\nu\beta\beta$ decay amplitude($M_{ee}$) is obtained to lie in the range ($0.04-0.06$)eV. Future $0\nu\beta\beta$ decay experiments such as NEXT and nEXO shall pose crucial test for the model. Both $CP$ conserving and $CP$ violating solutions are allowed in the model. Interesting correlations are obtained, specially, between Yukawa couplings of modular weight 2 and complex modulus $\tau$. Contrary to $Y_2^{2}$ and $Y_3^{2}$, the Yukawa coupling $Y_1^{2}$ is found to be insensitive to $\tau$ and thus to $CP$ violation because complex modulus $\tau$ is the only source of $CP$ violation in the model. We, also, investigate the prediction of the model for BAU. The model exhibit consistent explanation of BAU for right-handed Majorana neutrino mass scale in the range ($(1-5)\times10^{13}$)GeV.}
\begin{document} 
\maketitle
\flushbottom
\section{Introduction}
\label{sec:1}
\noindent The discovery of Higgs Boson at Large Hadron collider(LHC) has validated the standard model(SM) of particle physics as low energy effective theory. Although, SM explains the interactions of fundamental particles, mass generation  through  Higgs mechanism but flavour structure (masses and mixing) of fermions is unknown. There are some of mysteries/puzzles which still remain unanswered : origin of neutrino masses, large favor mixing in lepton sector, matter-antimatter asymmetry, to name few. In SM, neutrino is massless as dictated by underlying symmetry $SU(3)_{C}\times SU(2)_{L}\times U(1)_{Y}$. The absence of right-handed neutrino in SM prohibits neutrino mass generation through Higgs mechanism. The flavor mixing in lepton sector is large as compared to quark sector which is not addressed by SM. Also, $CP$ violation in quark sector is insufficient to explain observed matter-antimatter asymmetry which calls for additional sources of $CP$ violation in the leptonic sector. In view of these unexplained observations, beyond standard model (BSM) physics has been explored in different dimensions.     
\noindent The quest of understanding the origin of matter-antimatter asymmetry in the Universe and  underlying dynamics of neutrino mass generation is long standing. The framework of model building serves the platform which can address the simultaneous explanation for neutrino mass generation and observed matter-antimatter asymmetry through leptogenesis. 
In the top-down approach, the new particle content including usual SM particles, obeys some higher symmetry which could be a new Abelian global symmetry, non-Abelian discrete symmetry  or some cyclic symmetry. Neutrino oscillation data hints that lepton mixing could have group theoretical origin. Discrete symmetries have been useful in understanding the flavor structure of fermions and have been explored at the length in the literature, for review see \cite{King:2013eh,Altarelli:2010gt,Smirnov:2013uba,Cai:2017jrq,Herrero-Garcia:2019czj}. Non-Abelian discrete symmetries on spontaneous breaking have domain wall problem where one has to deal with distinct degenerate vacua \cite{Riva:2010jm,Antusch:2013toa,King:2018fke}. Also, origin of non-Abelian symmetries is not exactly known, although some works suggest origin of non-Abelian symmetries from continuous symmetries which require fine tuning such as breaking of gauged $SO(3) \rightarrow A_4$ \cite{Berger:2009tt}.

\noindent Recently, Modular symmetries have gained much attention in addressing flavor issues due to its connection to more fundamental theory for example, string theory. String theory of two dimensional torus have geometrical modular symmetry. Compactification of heterotic  and D-brane models \cite{Lauer:1989ax,Lauer:1990tm,Lerche:1989cs,Cremades:2004wa,Kobayashi:2017dyu} leads to modular symmetry and flavor groups $S_3$, $S_4$, $A_4$ and $A_5$ as its finite congruence subgroups of Modular groups. Modular symmetries differ from usual flavor symmetries in the sense that Yukawa couplings are not free. Yukawa couplings comprises of modular forms which, also, transform like other matter fields under modular symmetry \cite{Cremades:2004wa,Kobayashi:2017dyu}. A complex modulus $\tau$ breaks the symmetry of the theory which restricts the inclusion of extra flavon fields having specific vacuum expectation value ($vev$) alignments. Yukawa couplings of different order depend on the complex modulus $\tau$. Seesaw mechanisms remains the integral part of finite modular group based models.  Finite modular groups have been explored in different dimensions e.g. with in type-I+II seesaw \cite{Wang:2019xbo}, Type-II seesaw \cite{Kobayashi:2019gtp}, inverse seesaw realization \cite{Nomura:2019xsb,Nomura:2020cog}, linear seesaw realization \cite{Nomura:2020opk,Behera:2020sfe} and scotogenic or radiative scenarios \cite{Okada:2019xqk,Nomura:2019lnr,Behera:2020lpd}.   
\noindent In literature, different phenomenological ansatz like texture zeros\cite{Frampton:2002yf,Desai:2002sz,Xing:2002ta,Verma:2020gpl}, hybrid textures\cite{Kaneko:2005yz,Dev:2009he,Goswami:2008uv}, scaling neutrino mass matrix \cite{Mohapatra:2006xy,Grimus:2004cj,Joshipura:2009fu}, magic neutrino mass matrix \cite{Lam:2006wy,Channey:2018cfj,Verma:2019uiu} have been proposed to understand the mechanism of neutrino mass generation. The flavor theoretic realization of these ansatze, based on non-Abelian discrete groups, rely on introducing one or more flavon fields and thus, suffers from flavon alignment perplexities. Also, the emergence of the possible higher dimensional operators reduces the predictability of the flavor model. Texture zeros have been realized using finite modular group in \cite{Lu:2019vgm,Zhang:2019ngf}. 

    \noindent In the present work we focus on scaling mass matrix which was firstly proposed in Ref.\cite{Mohapatra:2006xy}. We have shown that modular forms of even weights naturally lead to scaling in the neutrino mass matrix provided we astutely assign the representations and weights under the modular group. Scaling requires that the ratio of certain elements of neutrino mass matrix are equal. The stability of scaling ansatz against renormalization group running from some seesaw scale to low scale is a sublime feature. Apart from inverted hierarchical neutrino mass, scaling ansatz results in vanishing reactor angle which is phenomenologically disallowed.  Here, we propose theoretical origin of scaling ansatz and its breaking in a modular $A_4$ symmetric model to have consistent low energy phenomenology and observed BAU. In this work, we propose two possible dynamically different scenarios of scale-breaking in the neutrino mass matrix based on (i) dimension-5 Weinberg operator and (ii) implementing Type-II seesaw via scalar triplet Higgs superfields($\Delta,\bar{\Delta}$) in a supersymmetric framework.  SM particle content is enlarged by introducing two right-handed neutrinos, one scalar field $\phi$\cite{King:2020qaj} and one scalar triplet Higgs field($\Delta$). At Type-I seesaw level, the interplay of two right-handed neutrinos and scalar field $\phi$ yields scaling in the neutrino mass matrix with vanishing lightest mass eigenvalue and $\theta_{13}=0$\cite{Mohapatra:2006xy,Blum:2007qm}. The breaking patterns of both the scenarios are proportional Yukawa coupling of modular weight 10. 

\noindent This paper is organized as follows. In Sec. \ref{sec:2}, we briefly introduce the finite modular groups and construction of higher modular weight Yukawa couplings. In Sec. \ref{sec:3}, we discuss the model setup, in detail. Numerical analysis and important results are discussed in Sec. \ref{sec:4}. In Sec. \ref{sec:5}, we have discussed the scenario for successful leptogenesis. Finally, we conclude in  Sec. \ref{sec:6}

\section{Modular Symmetry}
\label{sec:2}
Modular functions are subclass of functions which are invariant under linear fractional transformations defined as 
\begin{equation}
\gamma : z\rightarrow\gamma(\tau) =\frac{a\tau +b}{c\tau+d}, 
\end{equation}
such that $ad-bc=1$, where $a$, $b$, $c$ and $d$ are integers($Z$) and $\tau$ is a complex number. Modular group($\Gamma$)  is defined as group of these linear fractional transformations acting on upper half complex plane. It is isomorphic to $PSL(2,Z)$(projective special linear group) of $2\times2$ matrices with unit determinant and integers($Z$) as its elements. Modular group is generated by matrices
\[ S= \left( \begin{array}{cc}
0 & 1 \\
-1 & 0
\end{array} \right);
\hspace{0.7cm}
T=\left( \begin{array}{cc}
1 & 1 \\
0 & 1
\end{array} \right), 
\]
satisfying the relations $S^2=I$ and $(ST)^3=I$. Generators act on complex number $\tau$ as
$$ S: \tau \rightarrow -\frac{1}{\tau},\hspace{0.7 cm} T: \tau \rightarrow \tau+1.$$
Series of groups, $\Gamma$(N), is defined as
\begin{equation}
\Gamma(N)=\left\{
 \begin{bmatrix}
    a & b\\
     c &d
  \end{bmatrix} \hspace{0.2cm}
  \in \hspace{0.2cm}  SL(2,Z),\hspace{0.2cm}\begin{bmatrix}
    a & b\\
     c &d
  \end{bmatrix}=\begin{bmatrix}
    1 & 0\\
     0 &1
  \end{bmatrix} (mod N) 
  \right\},    
\end{equation}
such that $\Gamma(1)=SL(2,Z)$. Since $\gamma$ and $-\gamma$ determines same linear transformations, $\Bar{\Gamma}(N)$ which gives distinct linear transformations such that $\Bar{\Gamma}=\Bar{\Gamma}(1)=\Gamma(1)/\{I,-I\}=PSL(2,Z)$. It is to be observed that $\Bar{\Gamma}(N)=\Gamma(N)/\{I,-I\}$ for N$\leq$ 2  and $\Bar{\Gamma}(N)=\Gamma(N)$ for $N > 2$. These are also called as infinite modular groups. Finite modular groups  are defined as quotient group $\Gamma_N$ = $\Bar{\Gamma}/\Bar{\Gamma}(N)$. Finite Modular groups $\Gamma_N$ are isomorphic to permutation group such that $\Gamma_{2} \simeq S_{3}$ \cite{Kobayashi:2018vbk,Okada:2019xqk}, $\Gamma_{3} \simeq A_{4}$ \cite{Nomura:2019jxj,Zhang:2019ngf}, $\Gamma_{4} \simeq S_{4}$ \cite{King:2019vhv,Penedo:2018nmg,Kobayashi:2019mna} and $\Gamma_{5} \simeq A_{5}$ \cite{Novichkov:2018nkm,Ding:2019xna}.\\

\subsection{Modular Forms of different weights for $\Gamma_3$ $\simeq$ $A_4$ }

Here, we intend to describe mermorphic functions on upper half complex plane which remain invariant under all transformations in modular group such that $f(\gamma\tau)=f(\tau)$ ($\gamma$ being some transformation). The required invariance is very restrictive in nature. Instead, we focus on the mermorphic functions $f(\tau)$ so that upon transformations, $f(\gamma\tau)$  have same zeros and poles as that of $f(z)$. Modular forms $f(\tau)$ are holomorphic functions of weight 2$k$ and level $N$ with well defined transformation properties under the group $\Gamma(N)$ such that for $k\geq 0$,
\begin{equation}
  f(\gamma \tau)= (c\tau + d)^{2k} f(\tau),  
\end{equation}

 $$ \text{where,} \hspace{0.6cm} \gamma= \begin{bmatrix}
    a & b\\
     c & d
  \end{bmatrix}
  \hspace{0.2cm}\in \hspace{0.4cm} \Gamma(N).
$$
Modular forms are invariant under infinite modular group $\Gamma(N)$ up to factor $(c \tau + d)^{2k}$, but they indeed transforms under the finite modular group $\Gamma_{N}$. Modular forms spanned a linear space of finite dimension. It is possible to choose a basis in linear space, in which transformation is described by a unitary representation $\rho$ of $\Gamma_N$,
\begin{equation}
 f_{i}(\gamma \tau)= (c \tau +d)^{k} \rho_{ij}(\gamma) f_{j}(\tau); \hspace{0.5cm} \gamma \hspace{0.2cm}
  \in \hspace{0.2cm} \Gamma_{N}.   
\end{equation}
The transformations can be described by unitary representation. In fact, it is easy to observe that
  \begin{equation}
   f(\tau)\rightarrow e^{i \alpha} (c \tau +d)^{k} f(\tau).   
  \end{equation}  
  Further, for modular weight 2,
  $$ \frac{d}{d\tau}log f(\tau) \rightarrow (c\tau +d)^2 \frac{d}{d\tau}log f(\tau) + kc (c\tau+d), $$
  the inhomogeneous term, $kc (c\tau+d)$, should vanish for all modular weight. It means that invariance under finite modular group is preserved provided sum of modular weights should be zero i.e.  $\sum  k_{i} =0.$

\noindent Modular forms along with modular weights play crucial role in model building. For instance, if $N=3$, $\Gamma_{3}\simeq A_{4}$ serves as  non-linear realization of $A_4$ non-Abelian discrete symmetry . Linear space of modular forms having modular weight $2k$ and level $k$ have $2k+1$ dimension. It results in modular forms for different weights as:
  \begin{itemize}
      \item For $k=0$, there exist only one trivial modular form which is independent of $\tau$ and is constant.
      \item For $k=1$, there exist three modular forms which are linearly independent forming a triplet of $A_4$ having modular weight 2.
      \item Modular forms of higher weights can be constructed using the products of modular forms of weight 2.
      \end{itemize}
Dedekind eta-function is defined in upper half complex plane as
      \begin{equation}
          \eta(\tau)=q^{1/24} \sum_{n=1}^{\infty} (1-q^{n}), 
      \end{equation}
where $q=e^{i2\pi \tau}$ has important role in construction of modular forms. Modular forms of weight 2 are constructed using Dedekind $\eta$-function and its derivative as\cite{Feruglio:2017spp} 
    \begin{equation}
        \begin{aligned}
  Y_{1}^{2}(\tau) &= \frac{i}{2 \pi} \left[ 
 \frac{\eta^{'}(\tau/3)}{\eta(\tau/3)} + \frac{\eta^{'}((\tau+1)/3)}{\eta((\tau+1)/3)}+\frac{\eta^{'}((\tau+2)/3)}{\eta((\tau+2)/3)}-27\frac{\eta^{'}(3\tau)}{\eta(3\tau)}\right],\\
  Y_{2}^{2}(\tau) &= \frac{-i}{\pi} \left[ 
 \frac{\eta^{'}(\tau/3)}{\eta(\tau/3)} + \omega^{2}\frac{\eta^{'}((\tau+1)/3)}{\eta((\tau+1)/3)}+ \omega \frac{\eta^{'}((\tau+2)/3)}{\eta((\tau+2)/3)}\right],\\
 Y_{3}^{2}(\tau) &= \frac{-i}{\pi} \left[ 
 \frac{\eta^{'}(\tau/3)}{\eta(\tau/3)} + \omega \frac{\eta^{'}((\tau+1)/3)}{\eta((\tau+1)/3)}+ \omega^{2} \frac{\eta^{'}((\tau+2)/3)}{\eta((\tau+2)/3)}\right],
\end{aligned}
    \end{equation}  
where $\omega=e^{i2\pi/3}$. Modular forms satisfy the constraint
 \begin{equation}
   Y_{2}^{2}+2 Y_{1}^{2} Y_{3}^{2} =0.  
 \end{equation}
 The modular forms constructed above are arranged as $A_4$ triplet 
 $$ Y= \begin{pmatrix}
	Y_{1}^{2} \\
	Y_{2}^{2} \\
	Y_{3}^{2} \\
	\end{pmatrix},$$
	where $Y_{1}^2$, $Y_{2}^2$ and $Y_{3}^2$ are components of triplet with modular weight 2. Modular forms of higher weights can be constructed using modular forms of weight 2, as follows
\[
Y_{1}^{4}=\left((Y_{1}^{2})^2+2 Y_{2}^{2} Y_{3}^{2}\right), \quad Y_{1^{\prime}}^{4}=\left((Y_{3}^{2})^2+2 Y_{1}^{2} Y_{2}^{2}\right), \quad Y_{3}^{4}=
\left(\begin{array}{l}
(Y_{1}^{2})^2-Y_{2}^{2} Y_{3}^{2} \\
(Y_{3}^{2})^2-Y_{1}^{2} Y_{2}^{2} \\
(Y_{2}^{2})^2-Y_{1}^{2} Y_{3}^{2}
\end{array}\right),
\]
\[
\begin{array}{l}
Y_{1}^{6}=(Y_{1}^{2})^3+(Y_{2}^{2})^3+(Y_{3}^{2})^3-3 Y_{1}^{2} Y_{2}^{2} Y_{3}^{2}, \\
Y_{3,1}^{6}=\left((Y_{1}^{2})^2+2 Y_{2}^{2} Y_{3}^{2}\right)
\left(\begin{array}{l}
Y_{1}^{2} \\
Y_{2}^{2} \\
Y_{3}^{2}
\end{array}\right), 
\quad 
Y_{3,2}^{6}=\left((Y_{3}^{2})^2+2 Y_{1}^{2} Y_{2}^{2}\right)
\left(\begin{array}{l}
Y_{3}^{2} \\
Y_{1}^{2} \\
Y_{2}^{2}
\end{array}\right),
\end{array}
\]
and the total dimension is 7. For $k=8$, modular forms are
\[
Y_{1}^{8}=\left((Y_{1}^{2})^2+2 Y_{2}^{2} Y_{3}^{2}\right)^{2}, \quad Y_{1^{\prime}}^{(8)}=\left((Y_{1}^{2})^2+2 Y_{2}^{2} Y_{3}^{2}\right)\left((Y_{3}^{2})^2+2 Y_{1}^{2} Y_{2}^{2}\right), \quad Y_{1^{\prime \prime}}^{8}=\left((Y_{3}^{2})^2+2 Y_{1}^{2} Y_{2}^{2}\right)^{2},
\]
\[
Y_{3,1}^{8}=\left((Y_{1}^{2})^2+2 Y_{2}^{2} Y_{3}^{2}\right)\left(\begin{array}{l}
(Y_{1}^{2})^2-Y_{2}^{2} Y_{3}^{2} \\
(Y_{3}^{2})^2-Y_{1}^{2} Y_{2}^{2} \\
(Y_{2}^{2})^2-Y_{1}^{2} Y_{3}^{2}
\end{array}\right), \quad Y_{3,2}^{8}=\left((Y_{3}^{2})^2+2 Y_{1}^{2} Y_{2}^{2}\right)\left(\begin{array}{l}
(Y_{2}^{2})^2-Y_{1}^{2} Y_{3}^{2} \\
(Y_{1}^{2})^2-Y_{2}^{2} Y_{3}^{2} \\
(Y_{3}^{2})^2-Y_{1}^{2} Y_{2}^{2}
\end{array}\right),
\]
corresponding to a total dimension of 9. Also, Yukawa couplings of modular weight 10 are \\
\[
\begin{array}{l}
Y_{1}^{10}=\left((Y_{1}^{2})^2+2 Y_{2}^{2} Y_{3}^{2}\right)\left((Y_{1}^{2})^3+(Y_{2}^{2})^3+(Y_{3}^{2})^3-3 Y_{1}^{2} Y_{2}^{2} Y_{3}^{2}\right), \\
Y_{1^{\prime}}^{10}=\left((Y_{3}^{2})^2+2 Y_{1}^{2} Y_{2}^{2}\right)\left((Y_{1}^{2})^3+(Y_{2}^{2})^3+(Y_{3}^{2})^3-3 Y_{1}^{2} Y_{2}^{2} Y_{3}^{2}\right), \\
Y_{3,1}^{10}=\left((Y_{1}^{2})^2+2 Y_{2}^{2} Y_{3}^{2}\right)^{2}\left(\begin{array}{c}
Y_{1}^{2}\\
Y_{2}^{2} \\
Y_{3}^{2}
\end{array}\right), \quad Y_{3,2}^{10}=\left((Y_{3}^{2})^2+2 Y_{1}^{2} Y_{2}^{2}\right)^{2}\left(\begin{array}{c}
Y_{2}^{2} \\
Y_{3}^{2} \\
Y_{1}^{2}
\end{array}\right), \\
Y_{3,3}^{10}=\left((Y_{1}^{2})^2+2 Y_{2}^{2} Y_{3}^{2}\right)\left((Y_{3}^{2})^2+2 Y_{1}^{2} Y_{2}^{2}\right)\left(\begin{array}{c}
Y_{3}^{2} \\
Y_{1}^{2} \\
Y_{2}^{2}
\end{array}\right).
\end{array}
\]

\section{The Model}
\label{sec:3}
 
\noindent In this section, we construct the supersymmetric model in context of minimal Type-I+II seesaw scenario using $ A_{4}$ modular symmetry. We extend the usual SM content by two $CP$ conjugated superchiral neutrino fields $N^{c}_{1}$,  $N^{c}_{2}$ which are singlets under $A_4$ having SM representation $SU(2)_L$ $\times$ $U(1)_Y$ $\sim (1,0)$ and singlet `flavon' field (or weighton) $\phi$ $\sim$ (1,0). It may be noted here that $\phi$ do not compete with complex modulus $\tau$ for symmetry breaking as it being singlet does not have any vacuum expectation value ($vev$) alignment.  \\ 

\noindent $A_4$ irreducible representations are assigned such that charged lepton mass matrix and right-handed Majorana neutrino mass matrix obtained are diagonal. Lepton doublets $D_{iL}$ and $CP$ conjugated right-handed charged lepton superfields $e^{c}_{iR}$ are transforming as trivial or non-trivial $A_4$ singlets, each having modular weight -5. Higgs doublet superfields $H_u$, $H_d$ with hypercharges 1/2, -1/2 are assigned singlet representations of $A_4$ with modular weights 0, as shown in Table \ref{tab1}. 
The right-handed superfields $N^{c}_{1}$ ($A_4$ $\sim$ 1),  $N^{c}_{2}$ ($A_4$ $\sim$ $1'$) are assigned modular weights -1 and -3 resulting in diagonal right-handed Majorana mass matrix with scalar singlet field $\phi$ which is $A_4$  trivial singlet having modular weight equal to -2. \\

\noindent In order to write invariant superpotential, at tree level, for the field content considered in the model, we employ the Yukawa couplings($Y_m^n$) having even modular weights ($n= 4,6,8,10$ and $m=1, 1', 1''$) such that sum of modular weight is zero for each invariant term(Table \ref{tab2}). The product rules of $A_4$ modular symmetry discussed in the previous section are used to write higher order Yukawa couplings.\\

\begin{table}[t]
		\centering
		\begin{tabular}{cccccccccccc}
			Symmetry & ${D}_{eL}$  & ${D}_{\mu L}$ & ${D}_{\tau L}$& $e^{c}_{R}$ & $\mu^{c}_{R}$ & $\tau^{c}_{R}$ & $N^{c}_1$ & $N^{c}_2$ &$H_u$ & $H_d$   & $\phi$ \\ \hline
			$SU(2)_L$    &     2 &     2 &     2      &    1      &   1     &    1       & 1 & 1 &   2        &    2    &1  \\ \hline
			$U(1)_Y$    &     -$\frac{1}{2}$ & -$\frac{1}{2}$ &  -$\frac{1}{2}$ &   1      &   1     &    1       & 0 & 0 &   $\frac{1}{2}$        &    -$\frac{1}{2}$  & 0  \\  \hline
			$A_{4}$ &      1 &     1$'$ &     1$''$    &      1    &     1$''$      &     1$'$       & 1 & 1$'$  & 1          &    1    &1  \\ \hline
			-$k_{I}$ & -5 & -5 &-5 &-1 & -1 & -1 & -1& -3 & 0 & 0 & -2  \\ 
		\end{tabular}
		\caption{\label{tab1} Field content of the model and charge assignments under $SU(2)_L$, $U(1)_Y$, A$_{4}$ including modular weights.} 
	\end{table}
	\begin{table}[t]
\begin{center}
\begin{tabular}{ccccccccc}
 $Y^{n}_{m}$&  $Y^{4}_{1}$  & $Y^{4}_{1'}$  &  $Y^{6}_{1}$ &  $Y^{8}_{1}$ &  $Y^{8}_{1'}$  &  $Y^{8}_{1''}$&  $Y^{10}_{10}$ &  $Y^{10}_{1'}$ \\
 \hline
 A$_4$&1& 1$'$ & 1  & 1&1$'$  &  1$''$ &1 & 1$'$  \\
 \hline
	-$k_{I}$ & 4 & 4 & 6 & 8 & 8 & 8 & 10 & 10
\end{tabular}
\end{center}
\caption{\label{tab2} Transformation of Yukawa couplings of higher order under A$_{4}$ modular symmetry.}
\end{table}
\noindent The superpotential relevant for charged lepton masses reads as
\begin{equation} \label{charged}
W_{l}= \alpha'(D_{eL} H_{d} e^{c}_{R} Y^{6}_{1}) +\beta'(D_{\mu L} H_{d} \mu^{c}_{R} Y^{6}_{1})+\gamma'(D_{\tau L} H_{d} \tau^{c}_{R} Y^{6}_{1}),
\end{equation}
where $\alpha'$, $\beta'$ and $\gamma'$ are coupling constants and $Y_1^6$ is singlet Yukawa coupling of modular weight 6. \\
\noindent The vacuum expectation values of Higgs superfields, $H_u$ and $H_d$, are $v_{u}/ \sqrt{2}$ and $v_{d}/\sqrt{2}$, respectively. These are connected to the SM Higgs $vev$, as $v_{H}= \frac{1}{2}\sqrt{v_{u}^2 +v_{d}^2}$ and ratio of Higgs superfields is $\tan \beta=\frac{v_{u}}{v_{d}} $. Thus, charged lepton mass matrix takes the diagonal form as   \begin{equation} \label{charged_M}
	M_{l}=\frac{v_d}{\sqrt{2}}
	{\begin{pmatrix}
		\alpha' Y_{1}^{6}& 0&0 \\
		0  & \beta' Y_{1}^{6}& 0 \\
		0& 0 & \gamma' Y_{1}^{6}  
		\end{pmatrix}}.
	\end{equation}
	\noindent For Type-I seesaw mechanism, we have assigned the charges for the superfields in such a way to have diagonal right-handed neutrino  mass matrix. The superpotential relevant to neutrino mass generation through Type-I seesaw mechanism is 
 \begin{eqnarray} \label{neutrino}
\nonumber
W^{I}_{\nu}= && g_{1}(D_{eL} H_{u}  N^{c}_{1} Y^{6}_{1}) +g_{2}(D_{e L} H_{u}  N^{c}_{2} Y^{8}_{1''})+g_{3}(D_{\mu L} H_{u}N^{c}_{2} Y^{8}_{1'})+ g_{4}(D_{\tau L} H_{u}  N^{c}_{2} Y^{8}_{1}) \\  
 \nonumber
 && + y_{\phi_{1}}\phi N^{c}_{1} N^{c}_{1} Y^{4}_{1} + y_{\phi_{2}}\phi N^{c}_{2} N^{c}_{2} Y^{8}_{1'}, \\ 
\end{eqnarray}
where $g_{i}$ ($i=1,2,3,4$) are coupling constants. Assuming the $vev$ of scalar field $\phi$ to be $v_{\phi}$, the Dirac and right-handed Majorana neutrino mass matrices are given by
	\begin{equation} \label{dirac}
	m_{D}=\frac{v_{u}}{\sqrt{2}}
	{\begin{pmatrix}
		g_{1} Y^{6}_{1}& g_{2} Y^{8}_{1''}\\
		0 &g_{3} Y^{8}_{1'} \\
		0&g_{4} Y^{8}_{1}  
		\end{pmatrix}}
		\equiv {\begin{pmatrix}
		g'_{1} & g'_{2} \\
		0 &g'_{3}  \\
		0&g'_{4}  
		\end{pmatrix}},
	\end{equation}
\begin{equation} \label{right}
	m_{R}=
	{\begin{pmatrix}
	 y_{\phi_{1}}v_{\phi} Y^{4}_{1} & 0\\
		0 & y_{\phi_{2}}v_{\phi}Y^{8}_{1'}
		\end{pmatrix}} \equiv 	{\begin{pmatrix}
	 M_{1} Y^{4}_{1}& 0\\
		0 & M_{2}Y^{8}_{1'}
		\end{pmatrix}}
		\equiv 	{\begin{pmatrix}
	 M'_{1}& 0\\
		0 & M'_{2}
		\end{pmatrix}},
	\end{equation}
	respectively, where $y_{\phi_{i}}$($i=1,2$) are coupling constants. Type-I seesaw contribution to Majorana neutrino mass matrix is
	\begin{equation} \label{type1}
	m_{\nu_{1}} =-m_{D} m_{R}^{-1} m_{D}^{T} =  
	{\begin{pmatrix}
	\frac{g'^{2}_{1}}{M'_{1}} +\frac{g'^{2}_{2}}{M'_{2}}& \frac{g'_{2}g'_{3}}{M'_{2}}&\frac{g'_{3}g'_{4}}{M'_{2}}\\
	\frac{g'_{2}g'_{3}}{M'_{2}}&\frac{g'^{2}_{2}}{M'_{2}}& \frac{g'_{2}g'_{4}}{M'_{2}}\\
		\frac{g'_{3}g'_{4}}{M'_{2}}& \frac{g'_{2}g'_{4}}{M'_{2}}&\frac{g'^{2}_{4}}{M'_{2}}  
		\end{pmatrix}},
	\end{equation}
Majorana neutrino mass matrix in Eqn.(\ref{type1}) is scaled wherein column 3 is scaled with respect to column 2 by factor $\frac{g'_4}{g'_2}$.  The exact scaling in neutrino mass matrix provides $\theta_{13}=0$ and is thus, phenomenologically disallowed. In fact, we can have two dynamically distinct possibilities of scale-breaking in the neutrino mass matrix (Eqn. (\ref{type1})) \textit{viz.,} 
\begin{description}[align=left]
    \item [\textbf{Scenario 1:}] Introducing the dimension-5 Weinberg operator contributing to the scale-breaking and non-zero $\theta_{13}$ without requiring additional scalar field.
    \item [\textbf{Scenario 2:}] Alternatively, retaining the overall superpotential of the model to tree level dimension-4 where scale-breaking is accomplished via introducing scalar triplet Higgs superfields($\Delta,\bar{\Delta}$). Furthermore, it is to be noted that the terms like  $Y^{10}_{1}(D_{eL}H_{u})(D_{e L}H_{u})$ are at sub-leading mass dimension-5 and are suppressed(although allowed by the symmetries of the model)\cite{Asaka:2019vev}.
\end{description}
These two possibilities are discussed below.

\noindent \textbf{Scenario 1:} The dimension-5 Weinberg terms are not forbidden by the symmetry of the model and shall contribute to the overall structure of neutrino mass matrix. This possibility leads to the superpotential
\begin{eqnarray} \label{type-II}
\nonumber
W'^{II}_{\nu}= &&  \frac{1}{\Lambda}[a_{1} Y^{10}_{1} D_{e L} H_{u} H_{u} D_{e L} + a_{2} Y^{10}_{1'} (D_{e L} H_{u} H_{u} D_{\tau L}+D_{\tau L} H_{u} H_{u} D_{e L})\\
&&+a_{3} Y^{10}_{1'} D_{\mu L} H_{u} H_{u} D_{\mu L} 
+a_{4} Y^{10}_{1} (D_{\mu L}H_{u} H_{u} D_{\tau L}+D_{\tau L}H_{u} H_{u} D_{\mu L})],
\end{eqnarray}
where $a_{i}$ ($i=1,2,3,4$) are coupling constants and $\Lambda$ is scale of lepton number violation(LNV). After the SSB, the contribution to effective Majorana neutrino mass matrix is given by
\begin{equation} \label{Weinberg}
	m'_{\nu_{2}}=\frac{v_{u}^{2}}{\Lambda}
	{\begin{pmatrix}
		a_{1}Y_{1}^{10} &0& a_{2}Y_{1'}^{10}\\
      0&a_{3}Y_{1'}^{10}&a_{4}Y_{1}^{10}\\
		a_{2}Y_{1'}^{10}&a_{4}Y_{1}^{10} &0  
		\end{pmatrix}} \equiv 	{\begin{pmatrix}
	 a'_{1} & 0&a'_{2}\\
		0 & a'_{3}&a'_{4}\\
		a'_{2}&a'_{4}&0
		\end{pmatrix}}.
	\end{equation}
	In this case, the total effective Majorana neutrino mass matrix is 
  \begin{equation} \label{totalmnu}
m'_{\nu}=m_{\nu_1}+m'_{\nu_2}=
{\begin{pmatrix}
	\frac{g'^{2}_{1}}{M'_{1}} +\frac{g'^{2}_{2}}{M'_{2}}+a'_{1}& \frac{g'_{2}g'_{3}}{M'_{2}} &\frac{g'_{3}g'_{4}}{M'_{2}}+a'_{2}\\
	\frac{g'_{2}g'_{3}}{M'_{2}}&\frac{g'^{2}_{2}}{M'_{2}}+a'_{3}& \frac{g'_{2}g'_{4}}{M'_{2}}+a'_{4}\\
		\frac{g'_{3}g'_{4}}{M'_{2}}+a'_{2}& \frac{g'_{2}g'_{4}}{M'_{2}}+a'_{4}&\frac{g'^{2}_{4}}{M'_{2}} 
		\end{pmatrix}}.
	\end{equation}
	
\noindent \textbf{Scenario 2:} In order to achieve scale-breaking using triplet seesaw (Type-II seesaw), we have included one pair of scalar triplet Higgs superfields (equivalent to single scalar triplet Higgs field in non-supersymmetric version) $\Delta$ $\sim$ (3,1) and  $\Bar{\Delta}$ $\sim$ (3,-1) in $SU(2)_L$ $\times$ $U(1)_Y$ representation having vector notation $\Delta$ ( $\Delta^0$, $\Delta^{+}$,  $\Delta^{++}$), $\Bar{\Delta}$ ($\Bar{\Delta}^0$, $\Bar{\Delta}^{-}$, $\Bar{\Delta}^{--}$). The scalar triplets $\Delta$, $\Bar{\Delta}$ are assigned singlet representations of $A_4$ with modular weights 0. The Type-II seesaw contribution to neutrino masses emanates from the superpotential 
 \begin{eqnarray} \label{type-II}
\nonumber
W^{II}_{\nu}= && M \Delta \Bar{\Delta} +  \lambda_{1} H_{u} \Bar{\Delta} H_{u} + \lambda_{2} H_{d} \Delta H_{d} + K_{1} Y^{10}_{1} D_{e L} \Delta D_{e L} \\
\nonumber
&&+ K_{2} Y^{10}_{1'} (D_{e L} \Delta D_{\tau L}+D_{\tau L} \Delta D_{e L})+K_{3} Y^{10}_{1'} D_{\mu L} \Delta D_{\mu L}\\
&&+K_{4} Y^{10}_{1} (D_{\mu L} \Delta D_{\tau L}+D_{\tau L} \Delta D_{\mu L}), 
\end{eqnarray}
where $K_{i}$ ($i=1,2,3,4$) are coupling constants. Scalar triplet Higgs field acquires  $vev$ $v_{\Delta}$ resulting in Type-II contribution to effective Majorana neutrino mass matrix given by
  \begin{equation} \label{Type-II}
	m_{\nu_{2}}=v_\Delta
	{\begin{pmatrix}
		K_{1}Y_{1}^{10} &0& K_{2}Y_{1'}^{10}\\
      0&K_{3}Y_{1'}^{10}&K_{4}Y_{1}^{10}\\
		K_{2}Y_{1'}^{10}&K_{4}Y_{1}^{10} &0  
		\end{pmatrix}} \equiv 	{\begin{pmatrix}
	 K'_{1} & 0&K'_{2}\\
		0 & K'_{3}&K'_{4}\\
		K'_{2}&K'_{4}&0
		\end{pmatrix}}.
	\end{equation}
		The total neutrino mass matrix is 
  \begin{equation} \label{totalmnu2}
m_{\nu}=m_{\nu_1}+m_{\nu_2}=
{\begin{pmatrix}
	\frac{g'^{2}_{1}}{M'_{1}} +\frac{g'^{2}_{2}}{M'_{2}}+K'_{1}& \frac{g'_{2}g'_{3}}{M'_{2}} &\frac{g'_{3}g'_{4}}{M'_{2}}+K'_{2}\\
	\frac{g'_{2}g'_{3}}{M'_{2}}&\frac{g'^{2}_{2}}{M'_{2}}+K'_{3}& \frac{g'_{2}g'_{4}}{M'_{2}}+K'_{4}\\
		\frac{g'_{3}g'_{4}}{M'_{2}}+K'_{2}& \frac{g'_{2}g'_{4}}{M'_{2}}+K'_{4}&\frac{g'^{2}_{4}}{M'_{2}} 
		\end{pmatrix}}.
	\end{equation}	
It is interesting to note that the scale-breaking patterns in both(Eqns.(\ref{Weinberg}) and (\ref{Type-II})), otherwise dynamically different scenarios, are similar. This is due to the same charge assignments of superfields($\Delta,\bar{\Delta}$) and the Higgs superfield $H_u$ under modular $A_4$ symmetry(Table \ref{tab1}).

\noindent The model predictions, based on Eqn.(\ref{totalmnu2}), for neutrino mass, mixings and other derived quantities like $CP$ invariants ($J_{CP}, I_1, I_2$) and effective Majorana mass appearing in  neutrinoless double beta($0\nu \beta \beta$) decay, are discussed in the next section 

\section{Numerical Analysis}
\label{sec:4}
In charged-lepton basis, neutrino mixing matrix can be parameterized in terms of three mixing angles($\theta_{12},\theta_{23},\theta_{13}$), one Dirac-type($\delta$) and two Majorana-type$(\alpha, \beta)$ $CP$-violating phases viz.
\begin{equation}\label{UPMNS}
   U=\left(\begin{array}{ccc}
c_{12} c_{13} & s_{12} c_{13} & s_{13} e^{-i \delta} \\
-s_{12} c_{23}-c_{12} s_{23} s_{13} e^{i \delta} & c_{12} c_{23}-s_{12} s_{23} s_{13} e^{i \delta} & s_{23} c_{13} \\
s_{12} s_{23}-c_{12} c_{23} s_{13} e^{i \delta} & -c_{12} s_{23}-s_{12} c_{23} s_{13} e^{i \delta} & c_{23} c_{13}
\end{array}\right)\left(\begin{array}{ccc}
1 & 0 & 0 \\
0 & e^{i \frac{\alpha}{2}} & 0 \\
0 & 0 & e^{i \frac{\beta}{2}}
\end{array}\right) 
\end{equation}
where $c_{i j}=\cos\theta_{i j}$ and $s_{i j}=\sin\theta_{i j}$ ($i,j=1,2,3: i<j$). Using mixing matrix in Eqn.(\ref{UPMNS}), neutrino mixing angles are given by
\begin{equation} \label{angles}
\sin ^{2} \theta_{13}=\left|U_{e3}\right|^{2}, \quad \sin ^{2} \theta_{23}=\frac{\left|U_{\mu3}\right|^{2}}{1-\left|U_{e3}\right|^{2}}, \quad \sin ^{2} \theta_{12}=\frac{\left|U_{e2}\right|^{2}}{1-\left|U_{e3}\right|^{2}}.
\end{equation}

\begin{figure}
	\begin{center}
			{\epsfig{file=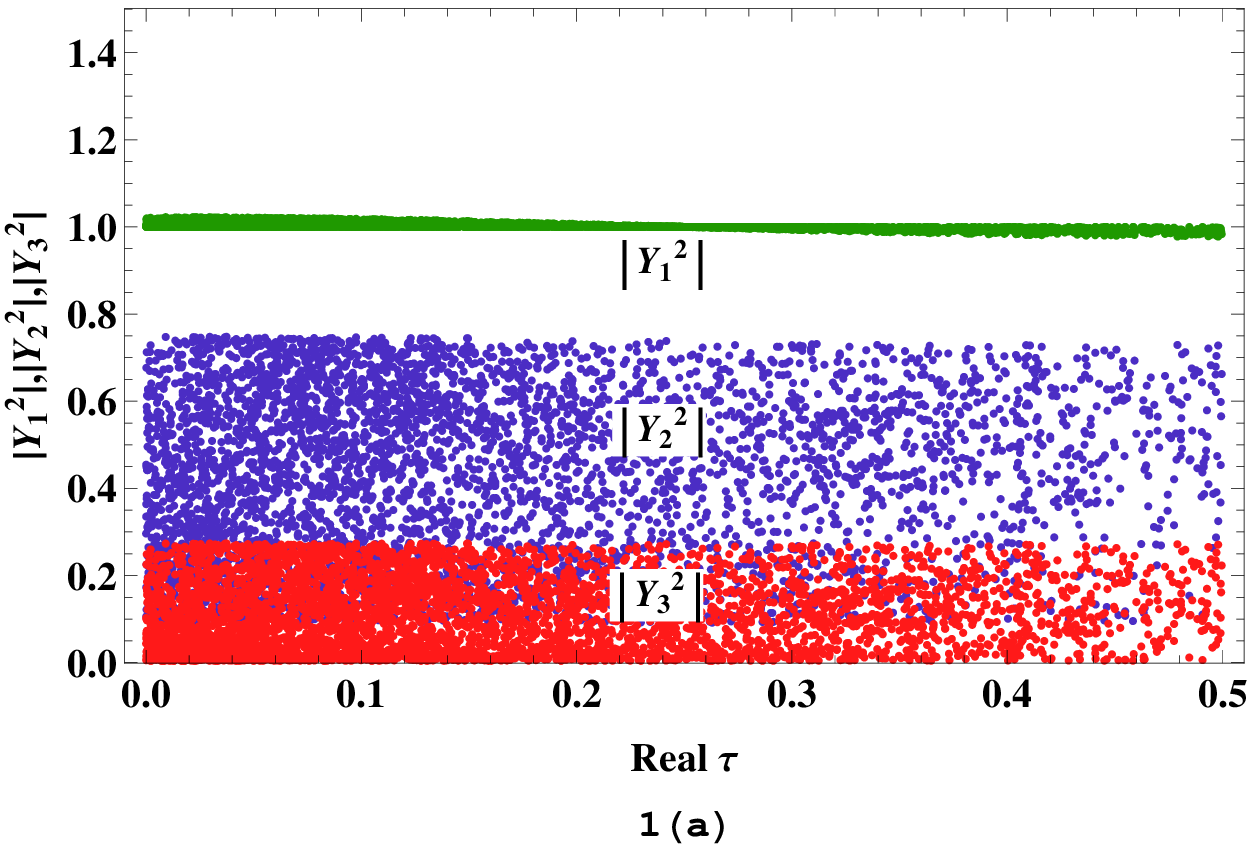,height=5.0cm,width=7.0cm}
				\epsfig{file=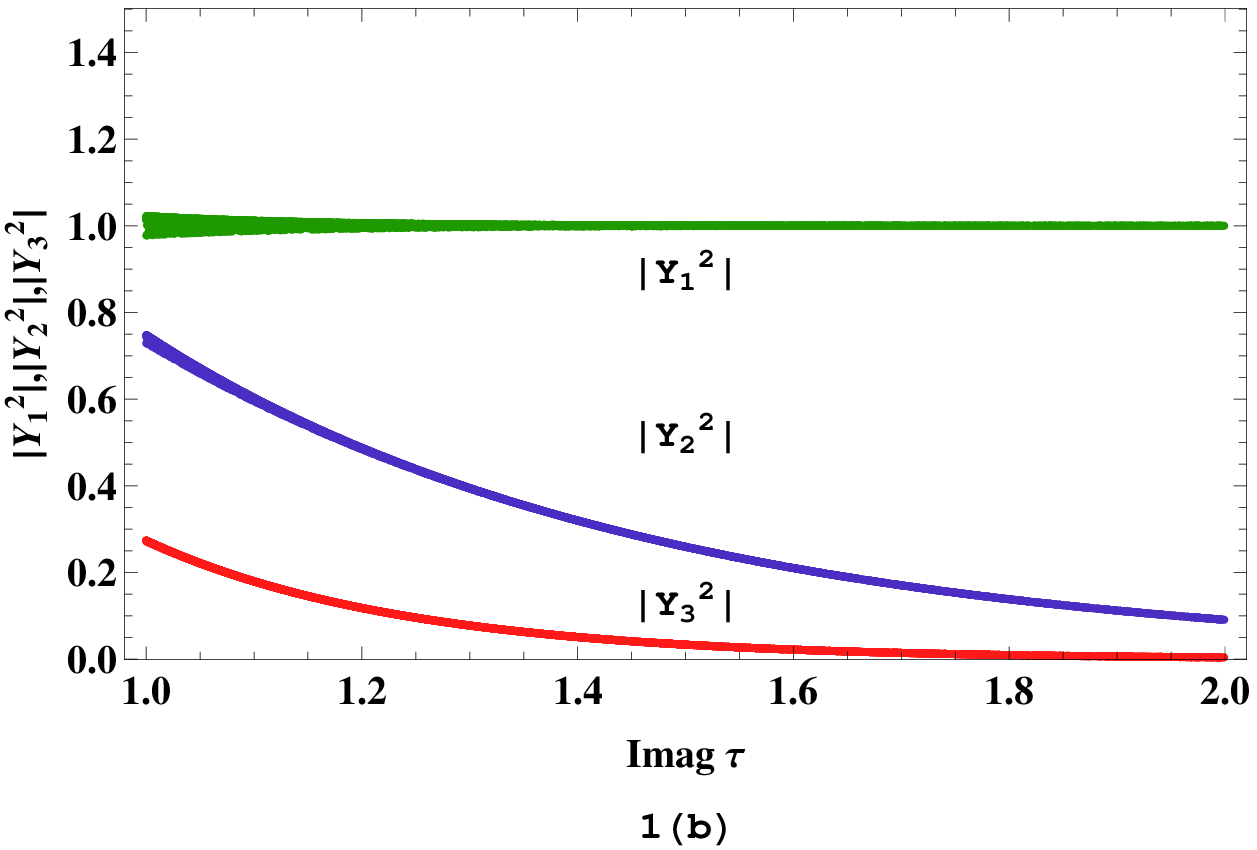,height=5.0cm,width=7.0cm}}
		\end{center}
\caption{\label{fig:1}  Variation of Yukawa couplings of modular weight 2 ($|Y_{1}^{2}|$, $|Y_{2}^{2}|$, $|Y_{3}^{2}|$) with real(Fig. 1(a)) and Imaginary part(Fig. 1(b)) of complex modulus $\tau$.}
\end{figure}
\begin{table}[h]
\begin{center}
\begin{tabular}{c|c|c}
\hline \hline 
Parameter & Best fit $\pm$ \( 1 \sigma \) range & \( 3 \sigma \) range  \\
\hline \multicolumn{2}{c} { Normal neutrino mass ordering \( \left(m_{1}<m_{2}<m_{3}\right) \)} \\
\hline \( \sin ^{2} \theta_{12} \) & $0.304^{+0.013}_{-0.012}$ & \( 0.269-0.343 \)  \\
\( \sin ^{2} \theta_{13} \) & $0.02221^{+0.00068}_{-0.00062}$ & \( 0.02034-0.02420 \) \\
\( \sin ^{2} \theta_{23} \) & $0.570^{+0.018}_{-0.024}$ & \( 0.407-0.618 \)  \\
\( \Delta m_{21}^{2}\left[10^{-5} \mathrm{eV}^{2}\right] \) & $7.42^{+0.21}_{-0.20}$& \( 6.82-8.04 \) \\
\( \Delta m_{31}^{2}\left[10^{-3} \mathrm{eV}^{2}\right] \) & $+2.541^{+0.028}_{-0.027}$ & \( +2.431-+2.598 \) \\
\hline \multicolumn{2}{c} { Inverted neutrino mass ordering \( \left(m_{3}<m_{1}<m_{2}\right) \)} \\
\hline \( \sin ^{2} \theta_{12} \) & $0.304^{+0.013}_{-0.012}$ & \( 0.269-0.343 \)\\
\( \sin ^{2} \theta_{13} \) & $0.02240^{+0.00062}_{-0.00062}$ & \( 0.02053-0.02436 \) \\
\( \sin ^{2} \theta_{23} \) & $0.575^{+0.017}_{-0.021}$& \( 0.411-0.621 \) \\
\( \Delta m_{21}^{2}\left[10^{-5} \mathrm{eV}^{2}\right] \) & $7.42^{+0.21}_{-0.20}$ & \( 6.82-8.04 \) \\
\( \Delta m_{32}^{2}\left[10^{-3} \mathrm{eV}^{2}\right] \) & $-2.497^{+0.028}_{-0.028}$ & \( -2.583--2.412 \)  \\
\hline \hline
\end{tabular}
\end{center}
\caption{\label{tab3}Experimental data of neutrino oscillation parameters from NuFIT 5.0 used in the numerical analysis\cite{Esteban:2020cvm}.}
\end{table}	
\noindent Also,  we can calculate the Jarlskog $CP$ invariant from matrix elements of $U$  as
\begin{equation} \label{jcp}
J_{C P}=\operatorname{Im}\left[U_{e 1} U_{\mu 2} U_{e 2}^{*} U_{\mu 1}^{*}\right]=s_{23} c_{23} s_{12} c_{12} s_{13} c_{13}^{2} \sin \delta,
\end{equation}
and other two $CP$ invariants \( I_{1} \) and \( I_{2} \) related to Majorana phases ($\alpha$, $\beta$) as
\begin{equation}\label{i1i2}
I_{1}=\operatorname{Im}\left[U_{e 1}^{*} U_{e 2}\right]=c_{12} s_{12} c_{13}^{2} \sin \left(\frac{\alpha}{2}\right), I_{2}=\operatorname{Im}\left[U_{e 1}^{*} U_{e 3}\right]=c_{12} s_{13} c_{13} \sin \left(\frac{\beta}{2}-\delta \right).    
\end{equation}
The effective mass for the neutrinoless double beta decay is given by
\begin{equation}
 M_{e e} =\left| m_{1} \cos ^{2} \theta_{12} \cos ^{2} \theta_{13} + m_{2} \sin ^{2} \theta_{12} \cos ^{2} \theta_{13} e^{i \alpha}+m_{3} \sin ^{2} \theta_{13} e^{i\left(\beta-2 \delta\right)}\right|, 
\end{equation}
The model prediction for $M_{ee}$ has been investigated in light of the $0\nu\beta\beta$ decay experiments like SuperNEMO \cite{Barabash:2012gc}, KamLAND-Zen \cite{KamLAND-Zen:2016pfg}, NEXT \cite{Granena:2009it,Gomez-Cadenas:2013lta}, nEXO \cite{Licciardi:2017oqg}.

\begin{figure}
	\begin{center}
			{\epsfig{file=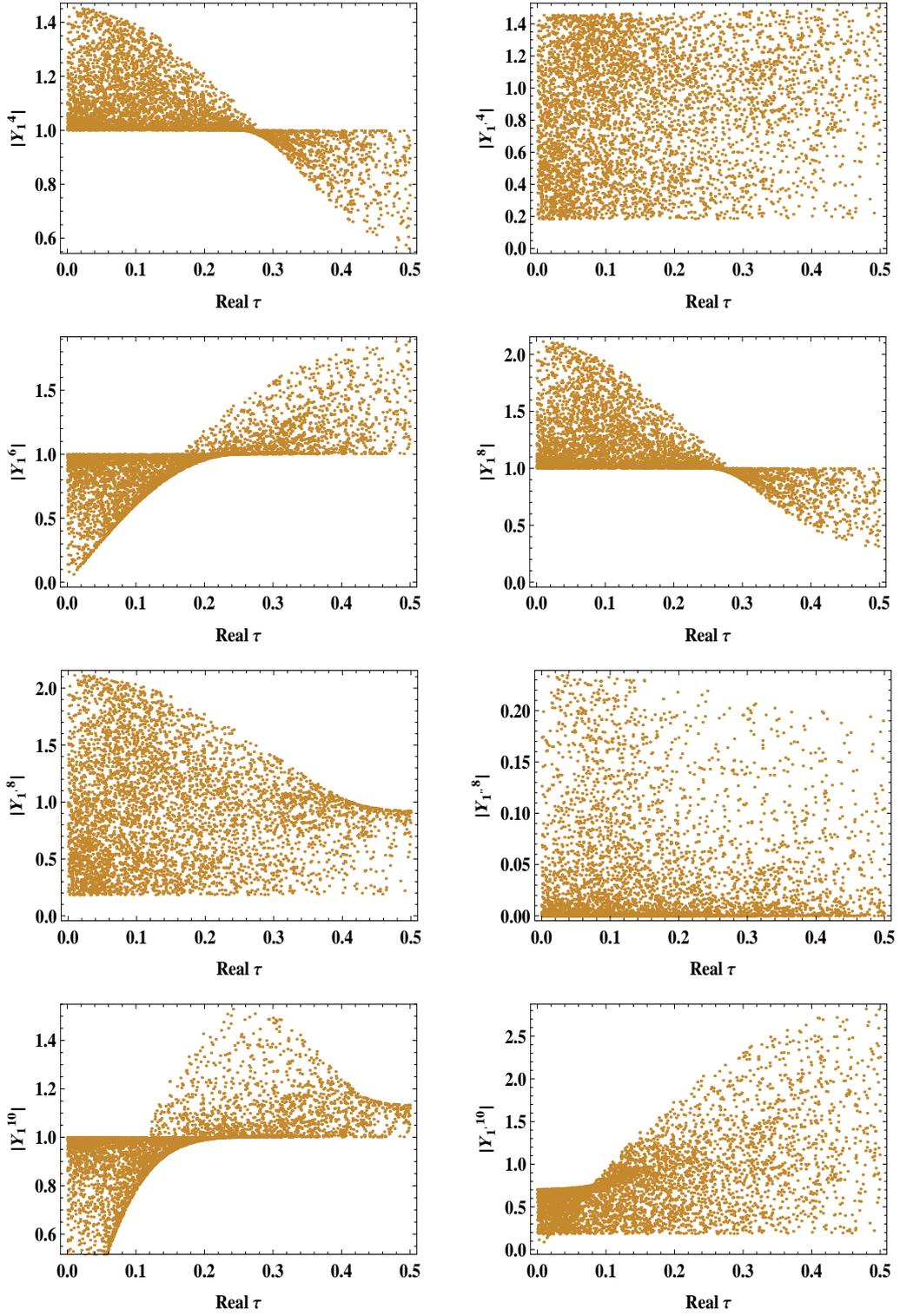,height=20.0cm,width=14.0cm}}
		\end{center}
		\caption{\label{fig:2}   Variation of Yukawa couplings of higher modular weight  with real part of complex modulus $\tau$.}
\end{figure}
\begin{figure}
	\begin{center}
			{\epsfig{file=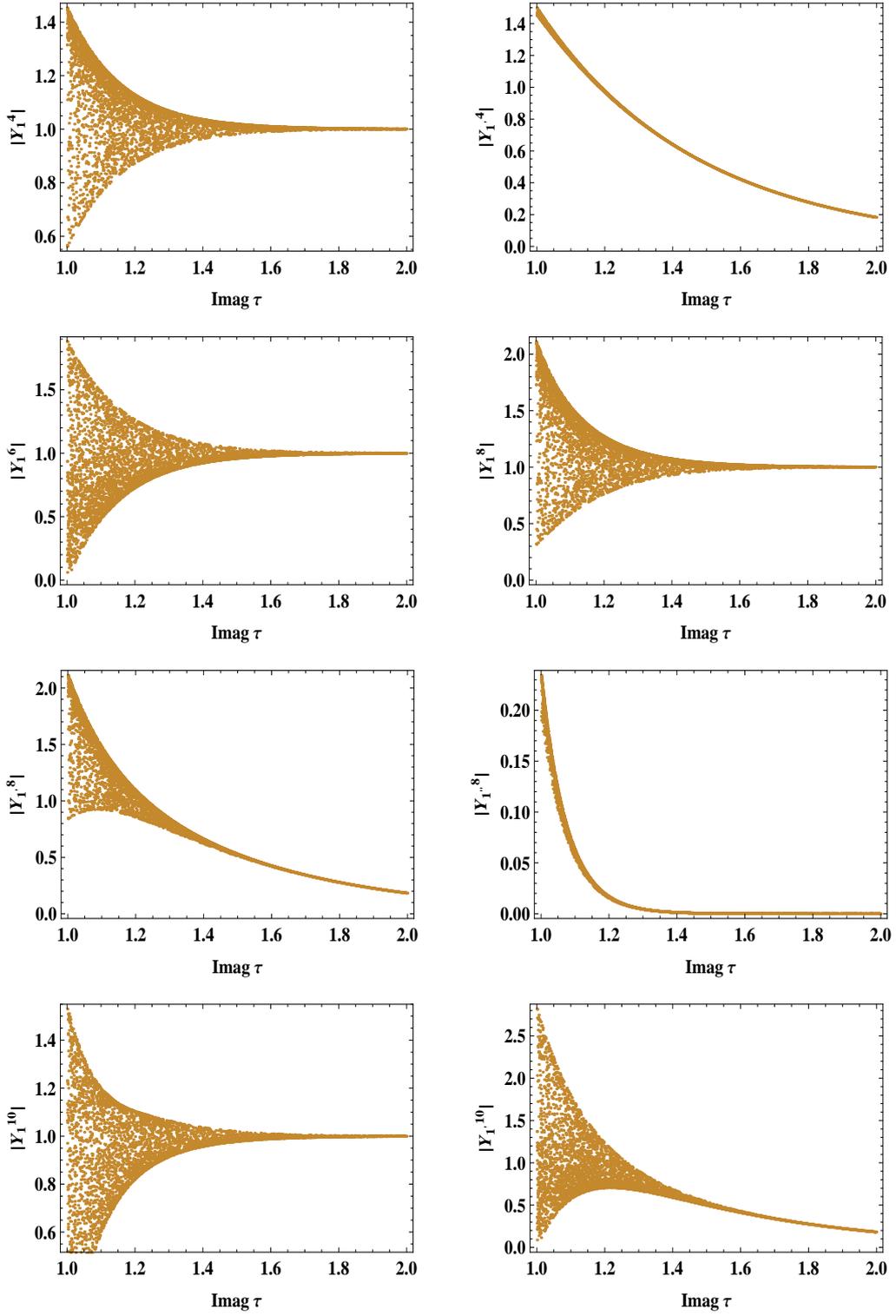,height=20.0cm,width=14.0cm}}
		\end{center}
\caption{\label{fig:3}   Variation of Yukawa couplings of higher modular weight with imaginary part of complex modulus $\tau$.}
\end{figure}

\begin{figure}
	\begin{center}
			{\epsfig{file=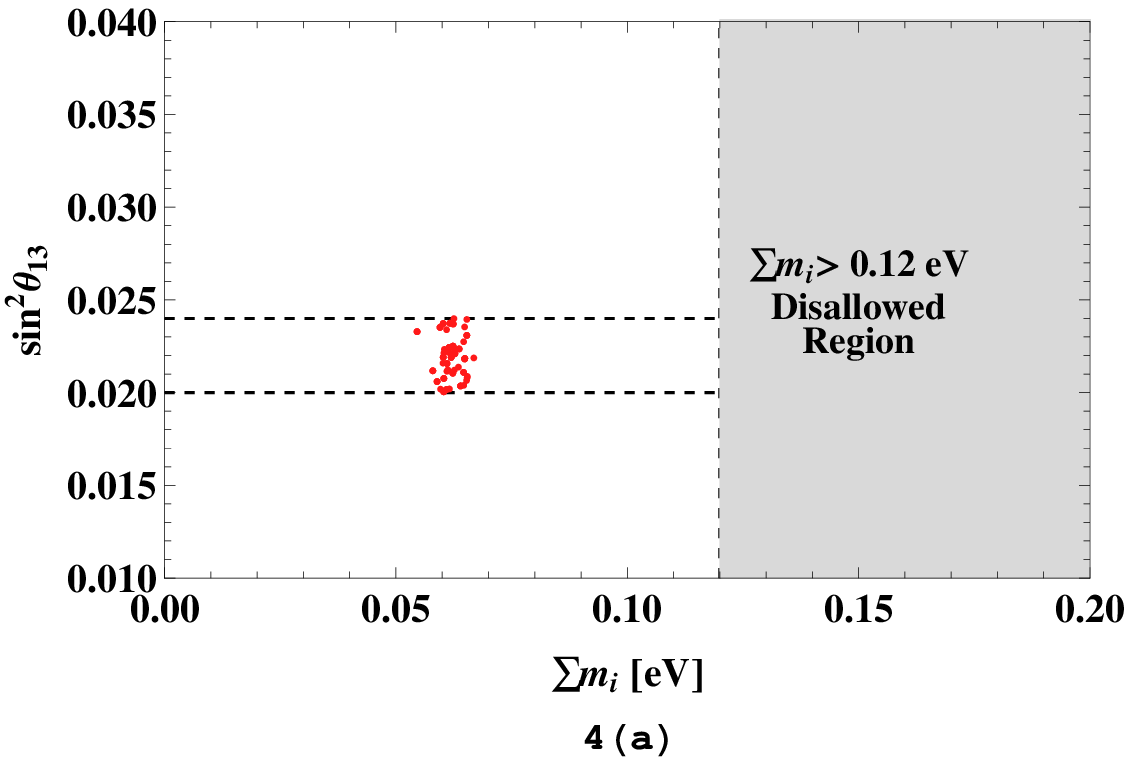,height=5.0cm,width=7.0cm}
				\epsfig{file=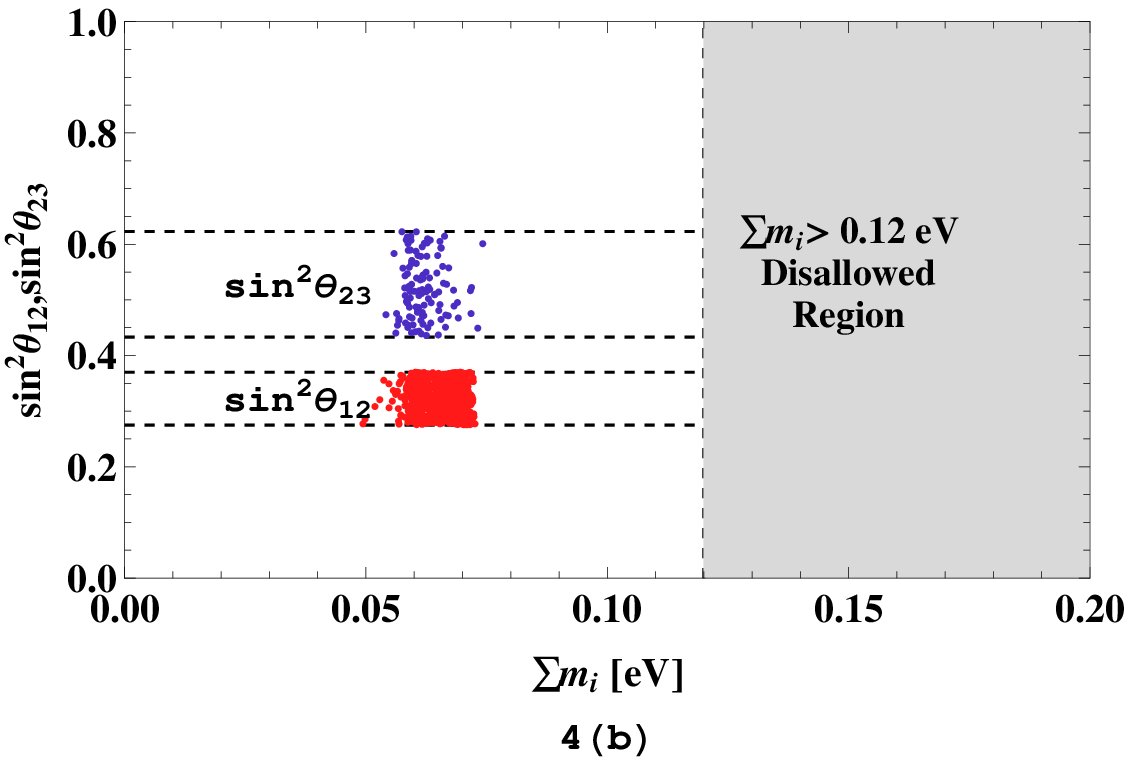,height=5.0cm,width=7.0cm}}
		\end{center}
\caption{\label{fig:4}  Variation of neutrino mixing angles with sum of neutrino masses $\sum m_{i}$  for Normal hierarchy. The grey shaded region is disallowed by cosmological bound on sum of neutrino masses\cite{Giusarma:2016phn,Aghanim:2018eyx}. Horizontal lines represent $3\sigma$ limits of respective mixing angle(Table \ref{tab3}).}
\end{figure}

\begin{figure}[ht!]
	\begin{center}
			{\epsfig{file=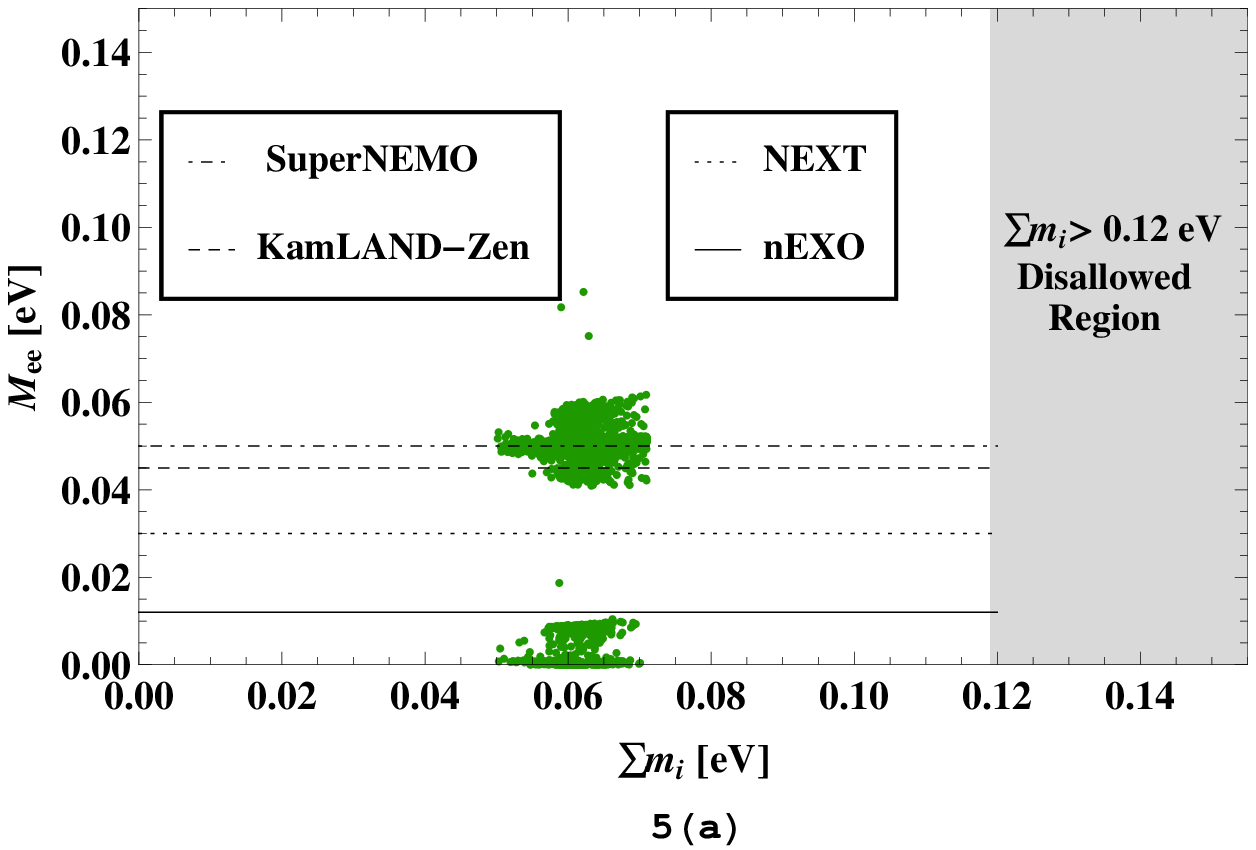,height=5.0cm,width=7.0cm}
				\epsfig{file=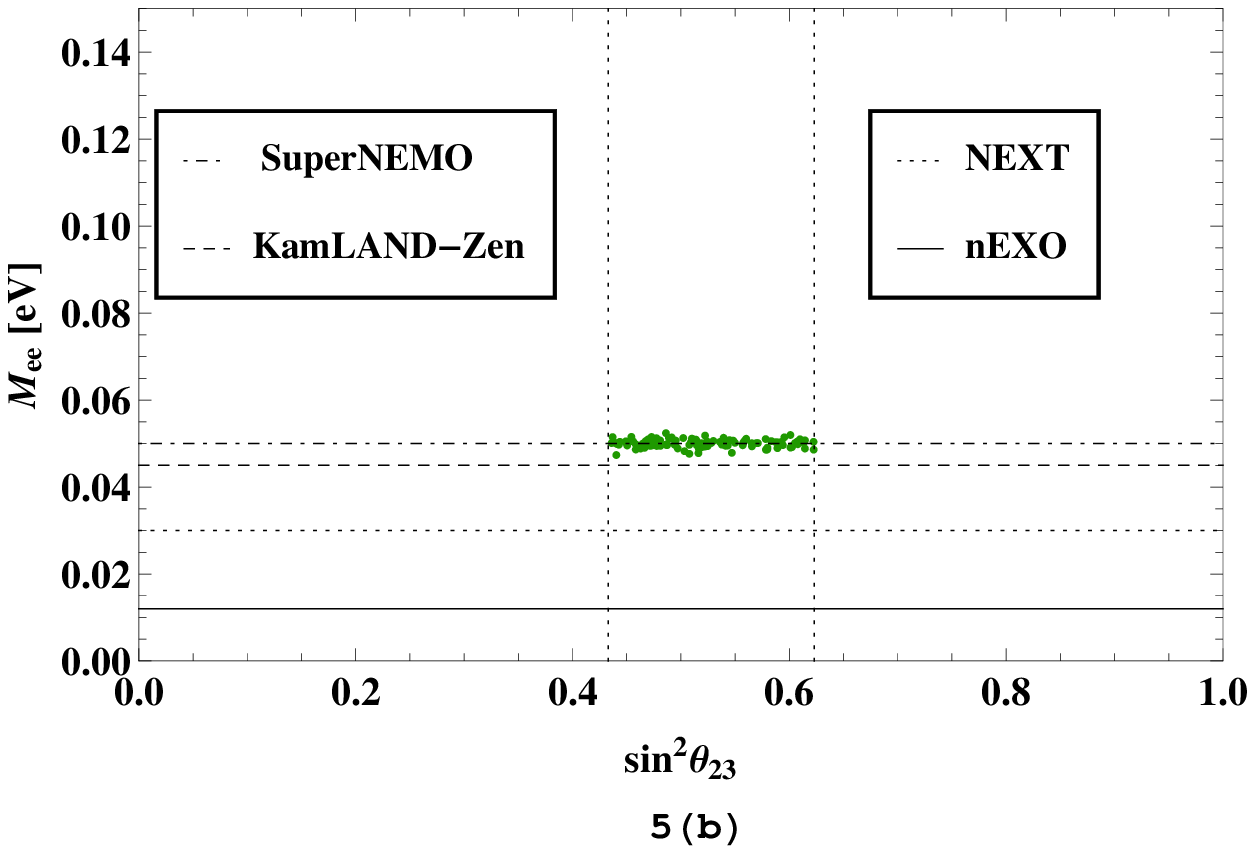,height=5.0cm,width=7.0cm}}
		\end{center}
\caption{\label{fig:5} Variation of Effective Majorana mass $|M_{ee}|$ with  sum of neutrino masses $\sum m_{i}$  for normal hierarchy (Fig. 5(a)) and with $\sin^{2}\theta_{23}$(Fig.5(b)). The horizontal lines are the sensitivity reach of $0\nu\beta\beta$ decay experiments. The grey shaded region is disallowed by cosmological bound on sum of neutrino masses\cite{Giusarma:2016phn,Aghanim:2018eyx}.}
\end{figure}

\begin{figure}[h]
 \begin{center}
 \epsfig{file=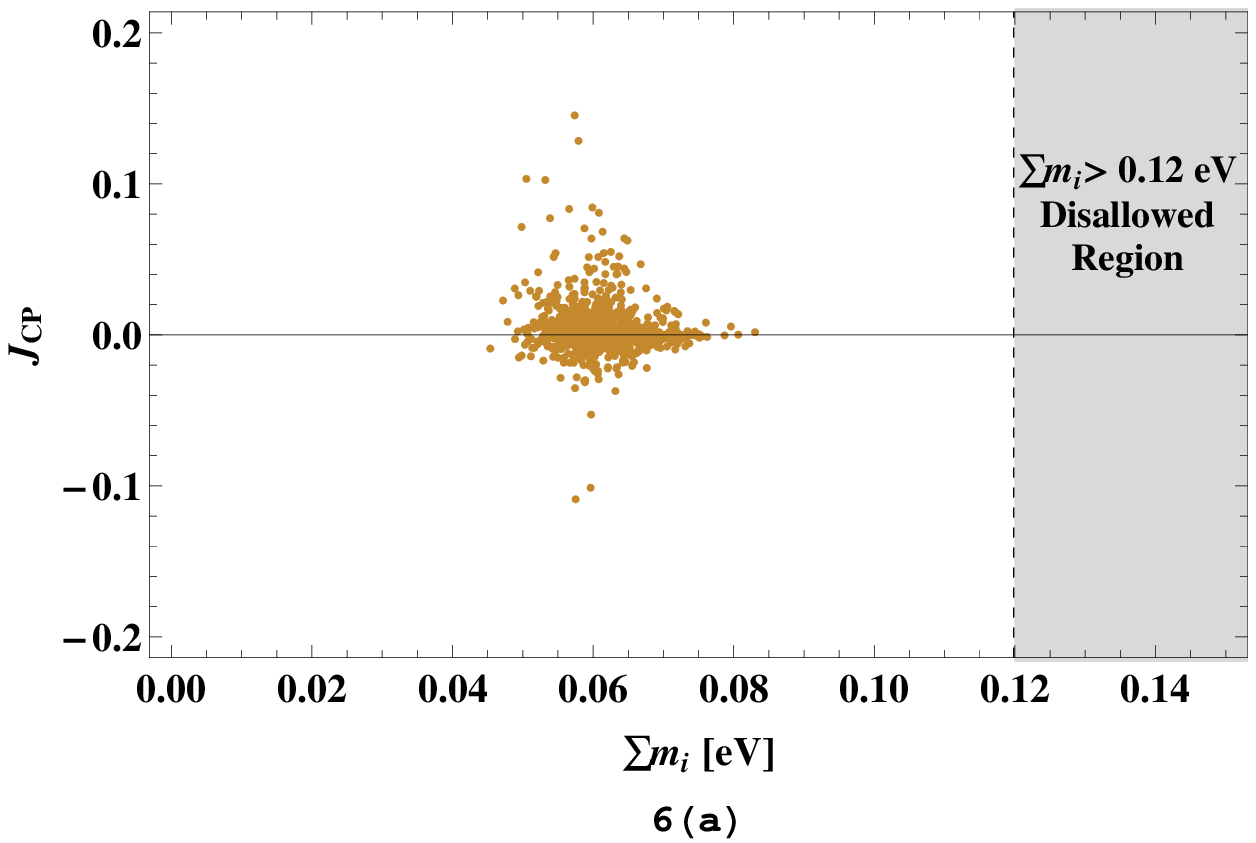,height=5.0cm,width=8.0cm}\\
 \vspace{0.5 cm}
{\epsfig{file=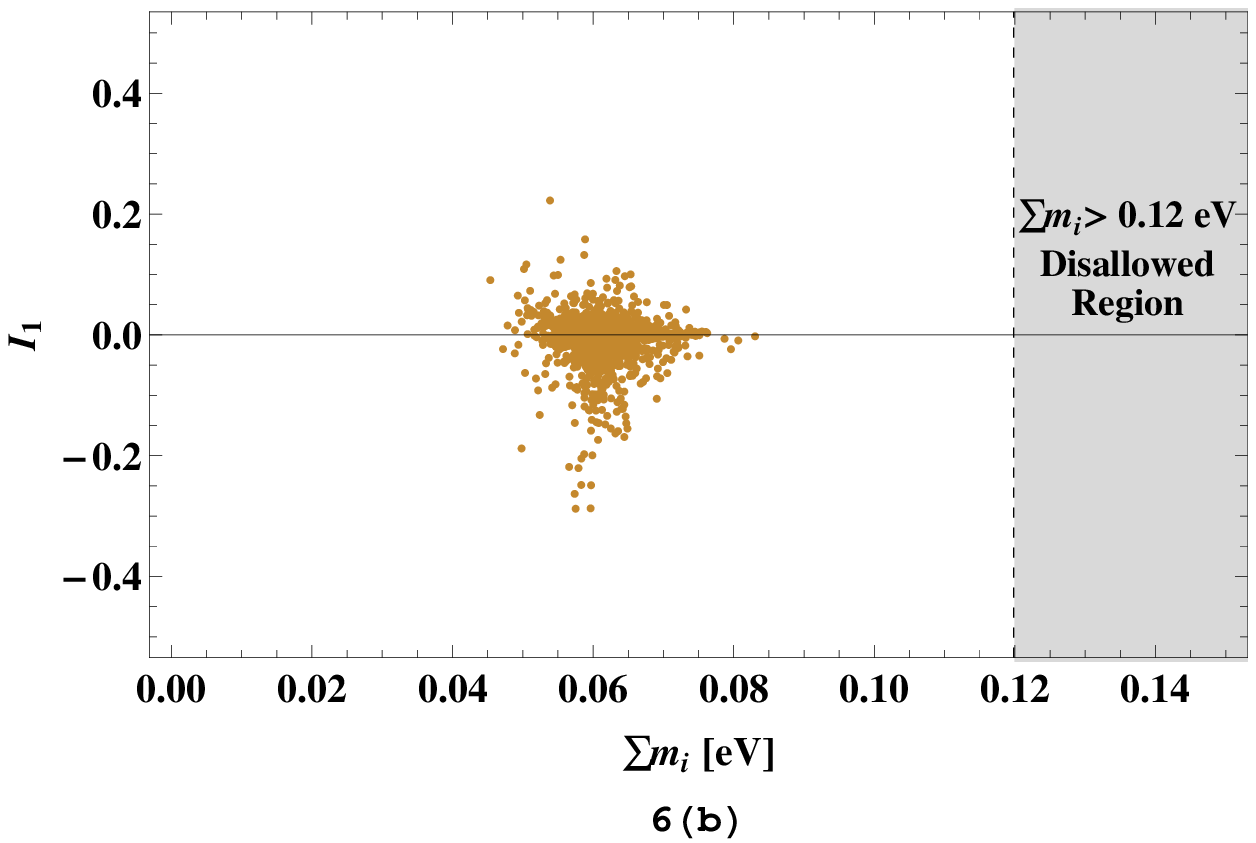,height=5.0cm,width=7.0cm},
\epsfig{file=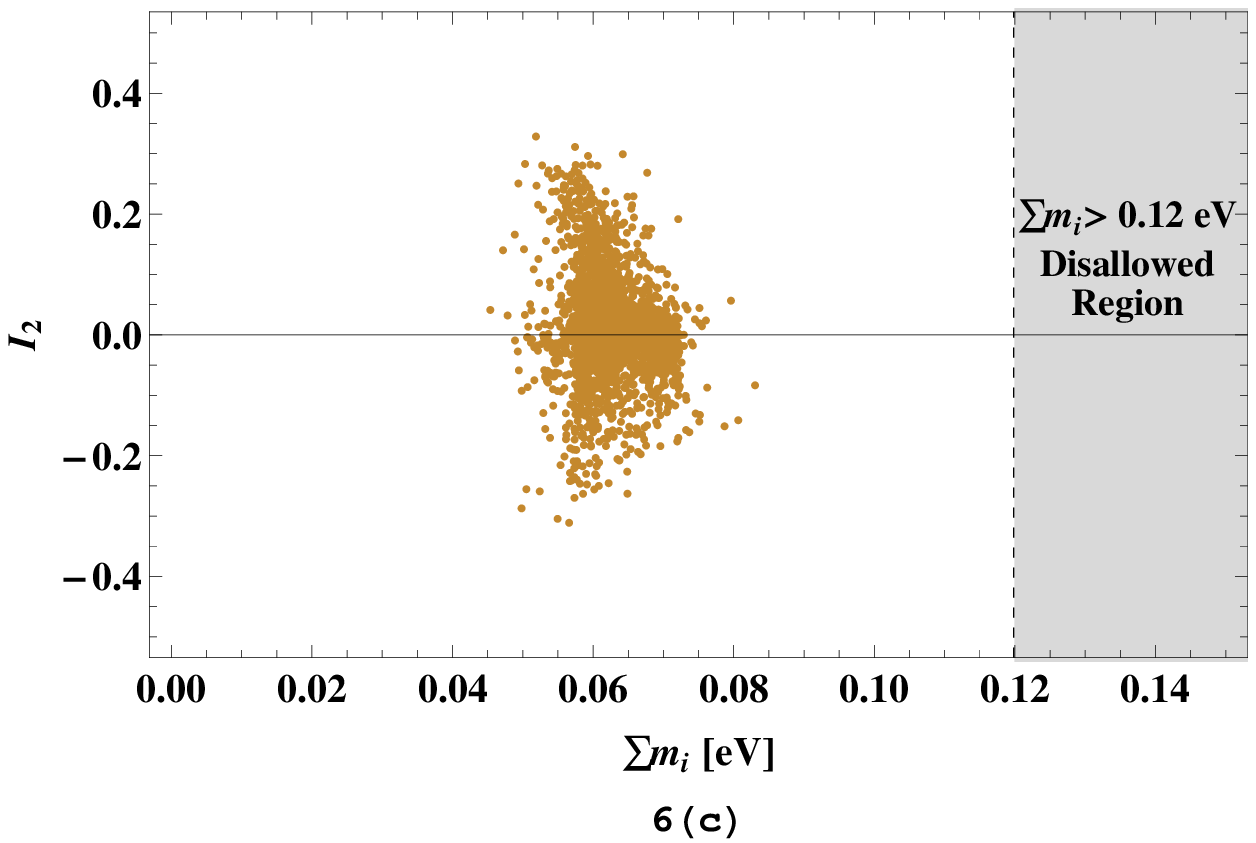,height=5.0cm,width=7.0cm}}
\end{center}
  \caption{\label{fig:6}Variation of  $CP$ invariants  $J_{CP}$,  $I_1$ and $I_2$  with sum of neutrino masses $\sum m_{i}$.} 
\end{figure}

\begin{figure}[ht!]
	\begin{center}
			{\epsfig{file=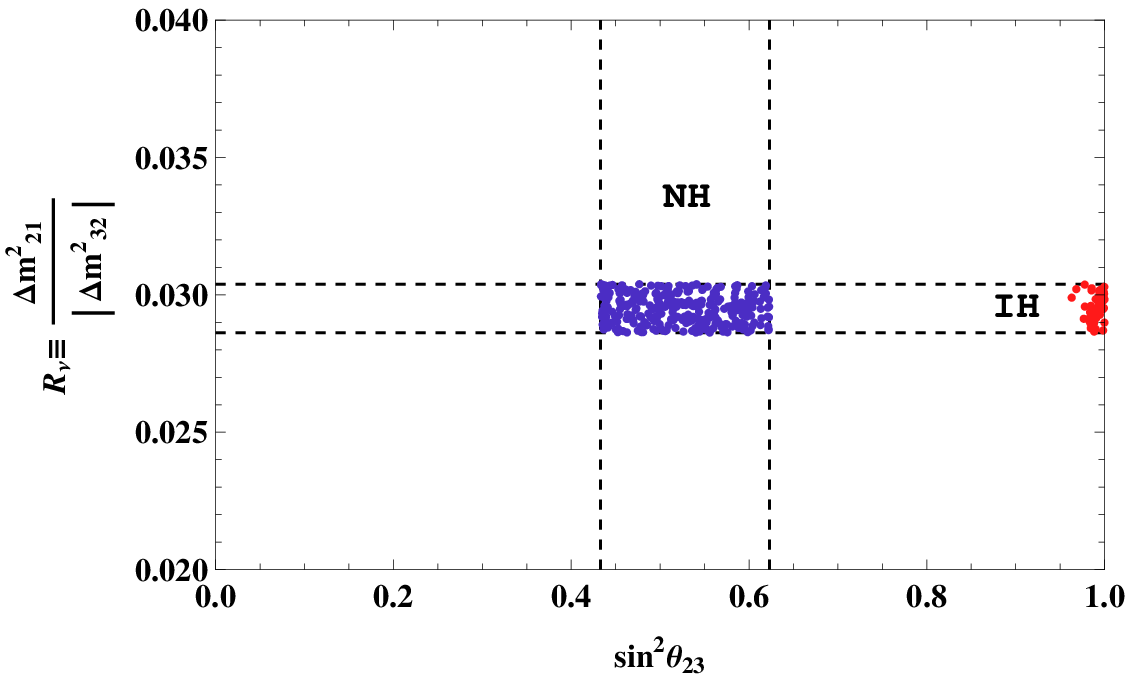,height=5.0cm,width=7.0cm}}
		\end{center}
\caption{\label{fig:7}  Allowed parameter space in ($\sin^{2}{\theta_{23}}$-$R_{\nu}$) plane for normal and inverted hierarchies. Horizontal(Vertical) lines represent $3\sigma$ experimental limits of $R_{\nu}$($\sin^{2}\theta_{23}$).}
\end{figure}

\noindent Using the relations derived above and obtained structure of effective Majorana neutrino mass matrix(Eqn.(\ref{totalmnu})), we carry out the numerical analysis. The model contains free real coupling constants $g_i, (i=1,2,3,4)$ and $K_i(i=1,2,3,4)$ which are varied randomly in the range $(0.01-1)$. Yukawa couplings are dependent on complex modulus $\tau$ whose real and imaginary parts are varied in the fundamental domain i.e. $|Re(\tau)| \leq 0.5$, $Im(\tau)> 0$, while $|\tau|>1$ \cite{Wang:2019xbo}. We have worked in the approximation in which $\tan \beta= 5$ \cite{Okada:2019uoy,Antusch:2013jca,Bjorkeroth:2015ora}. Also, vacuum expectation value($vev$) of Higgs fields ($v_{H}$) is $246$ GeV while Type-II scalar triplet field assumes $vev$ $\mathcal{O}(0.01)$eV \cite{Borah:2014bda}. The  lightest right-handed neutrino mass scale i.e. $M_1$ is varied randomly between $(1-5)\times10^{13}$ GeV whereas $M_2$ is varied in the range $ (1-5)\times10^{14}$ to ascertain the non-degenerate right-handed neutrino masses required for successful leptogenesis. 

\noindent We numerically diagonalize the neutrino mass matrix(Eqn.(\ref{totalmnu})) to obtain mass eigenvalues ($m_1,m_2,m_3$) from which model prediction for the mass-squared differences and mixing angles have been obtained. Using the experimental values of the mass-squared differences(Table \ref{tab3}) the allowed parameter space for mixing angles, $CP$ invariants and $M_{ee}$ have been obtained. The model satisfies the neutrino oscillation data for normal hierarchical neutrino mass spectrum. 
\noindent The Yukawa couplings of modular weight 2 are shown as a function of real and imaginary part of complex modulus $\tau$ in Fig.(\ref{fig:1}). The Yukawa couplings of higher modular weight can be constructed from the Yukawa couplings of modular weight 2. Also, Yukawa couplings having modular weights ($4,6,8,10$) are functions of complex modulus $\tau$ which have been used to write invariant superpotential. The variation of these Yukawa couplings with real and imaginary parts of complex modulus $\tau$ are shown in Fig.(\ref{fig:2}) and Fig.(\ref{fig:3}), respectively. Using the eigenvectors obtained from the diagonalization of Eqn.(\ref{totalmnu}) alongwith Eqn.(\ref{angles}), we obtain predictions for the mixing angles. In Fig.(\ref{fig:4}), we depict the correlation between sum of neutrinos mass eigenvalues ($\Sigma m_{i}$) and neutrino mixing angles ($\theta_{12}, \theta_{23}, \theta_{13}$). It  is evident from Fig.(\ref{fig:4}) that a very narrow range of sum of neutrino masses $(0.05-0.08)$eV is allowed, which is well below the current cosmological bound \cite{Giusarma:2016phn,Aghanim:2018eyx}. Also, the implication of the model for lepton number violating $0\nu\beta\beta$ decay is shown in Fig.(\ref{fig:5}). The allowed parameter space of effective Majorana mass $M_{ee}$ and sum of neutrino masses constitutes two regions degenerated in the range of $\Sigma m_{i} =  (0.05-0.08)$eV. The lower(upper) region corresponds to $M_{ee}$ in the range $(0-0.01)$eV($(0.04-0.06)$eV). The sensitivity reach of $0\nu\beta\beta$ decay experiments like SuperNEMO \cite{Barabash:2012gc}, KamLAND-Zen \cite{KamLAND-Zen:2016pfg}, NEXT \cite{Granena:2009it,Gomez-Cadenas:2013lta}, nEXO \cite{Licciardi:2017oqg} are, also, shown in the figure. It is evident from ($M_{ee}$-$\sin^{2}{\theta_{23}}$) plot(Fig.5(b)) that the lower region of $M_{ee}$ (see Fig.5(a)) is disallowed as it correspond to $\theta_{23}$ outside experimental range. Hence, model predicts the neutrinoless double beta decay amplitude to be with in the range ($(0.04-0.06)$eV) which is well within the sensitivity reach of $0\nu\beta\beta$ decay experiments. The prediction for the $CP$ invariants $J_{CP}$, $I_1$ and $I_2$ are shown in Fig.(\ref{fig:6}). 
In Fig.(\ref{fig:7}), we have shown prediction of the model for possible neutrino mass hierarchy. From the correlation between $R_{\nu}\equiv \frac{\Delta m_{21}^2}{|\Delta m_{32}^2|}$ and $\sin^{2} \theta_{23}$, it is evident that model predicts normal hierarchical neutrino masses(NH) because inverted hierarchy(IH) does not follow experimental bounds on $\sin^{2} \theta_{23}$. 

\noindent In the next section, we discuss leptogenesis framework in the current model setup to explain baryon asymmetry of the Universe.

\section{Leptogenesis}
\label{sec:5}
Within the framework of Type-I+II  seesaw, there can be contribution to baryon asymmetry from the decay of both heavy fields i.e. right-handed neutrino as well as scalar triplet. Also, it is necessary to obey the Sakharov's condition to observe successful leptogenesis \cite{Sakharov:1967dj}. For hierarchical spectrum of new heavy fields  such that  $M_{1}^{'}<<M_{2}^{'}$ and $M_{1}^{'}<<M_{\Delta}$, only right-handed decay mode will be dominant one. Here, lepton number violation and $CP$ asymmetry is generated by the decay of right-handed neutrino fields \cite{Domcke:2020quw}. This lepton number violation is then converted to baryon number violation via sphaleron processes with some efficiency factor ($K_{eff}$) dictating $CP$ asymmetry being washout, during conversion. The right-handed neutrino decay contributes to $CP$ asymmetry, $\epsilon_{N}$, through the interference of tree level and one loop decay processes (shown in Figs. 8(a) and 8(b)). In the approximation $M_{1}^{'}<<M_{2}^{'}$, the $CP$ asymmetry can be written as \cite{Antusch:2004xy,Hambye:2003ka}

	\begin{equation}\label{epsilonN}
	\epsilon_{N}=\frac{3 M'_1}{16 \pi v_{H}^{2}} \frac{Im[m_{D}^{\dagger}m_{\nu_1}m_{D}^{*}]_{11}}{(m_{D}^{\dagger}m_{D})_{11}}.
	\end{equation}
The contribution to $CP$ asymmetry from Type-II seesaw, at one loop, is through the decay of right-handed neutrino with scalar triplet mediating as virtual particle  as shown in Fig. 8(c). In the approximation, $M_{1}^{'}<<M_{\Delta}$, $CP$ asymmetry is given by \cite{Antusch:2004xy,Hambye:2003ka} 
	\begin{equation}\label{epsilonD}
	\epsilon_{\Delta}=\frac{3 M'_1}{16\pi v_{H}^{2}} \frac{Im[m_{D}^{\dagger}m_{\nu_2}m_{D}^{*}]_{11}}{(m_{D}^{\dagger}m_{D})_{11}}.
	\end{equation}
The total $CP$ asymmetry $\epsilon=\epsilon_{N_1}+\epsilon_{\Delta}$ is then converted into baryon asymmetry through sphaleron processes.  In the strong washout regime, efficiency factor is \cite{Buchmuller:2004tu} 
\begin{equation} \label{efficiency}
	K_{eff}=2\times10^{-2}\left(\frac{M'_1\times0.01eV}{(m_{D}^{\dagger}m_{D})_{11}}\right)^{1.1}.
	\end{equation}
	The out of equilibrium decay of right-handed neutrino requires that $\bar{m}<m^{*}$, where effective neutrino mass, $\bar{m} =\frac{(m_{D}^{\dagger}m_{D})_{11}}{M'_1}$ and equilibrium neutrino mass $m^{*}=\sqrt\frac{64g^{*}\pi^5}{45}.\frac{v_{H}^{2}}{M_{pl}}\approx 1.08\times10^{-3}$eV. The baryon asymmetry thus produced can be approximated as 
		\begin{equation}\label{eta}
	|\eta_B|\approx 0.96\times 10^{-2} \epsilon K_{eff}.
	\end{equation}
Using allowed model parameter space of $m_D$, $m_{\nu_1}$ and $m_{\nu_2}$ obtained in Sec.(\ref{sec:4}) we evaluated the $CP$ asymmetry with the help of Eqns. (\ref{epsilonN}) and (\ref{epsilonD})
\begin{figure} [t] 
		\includegraphics[scale=0.23]{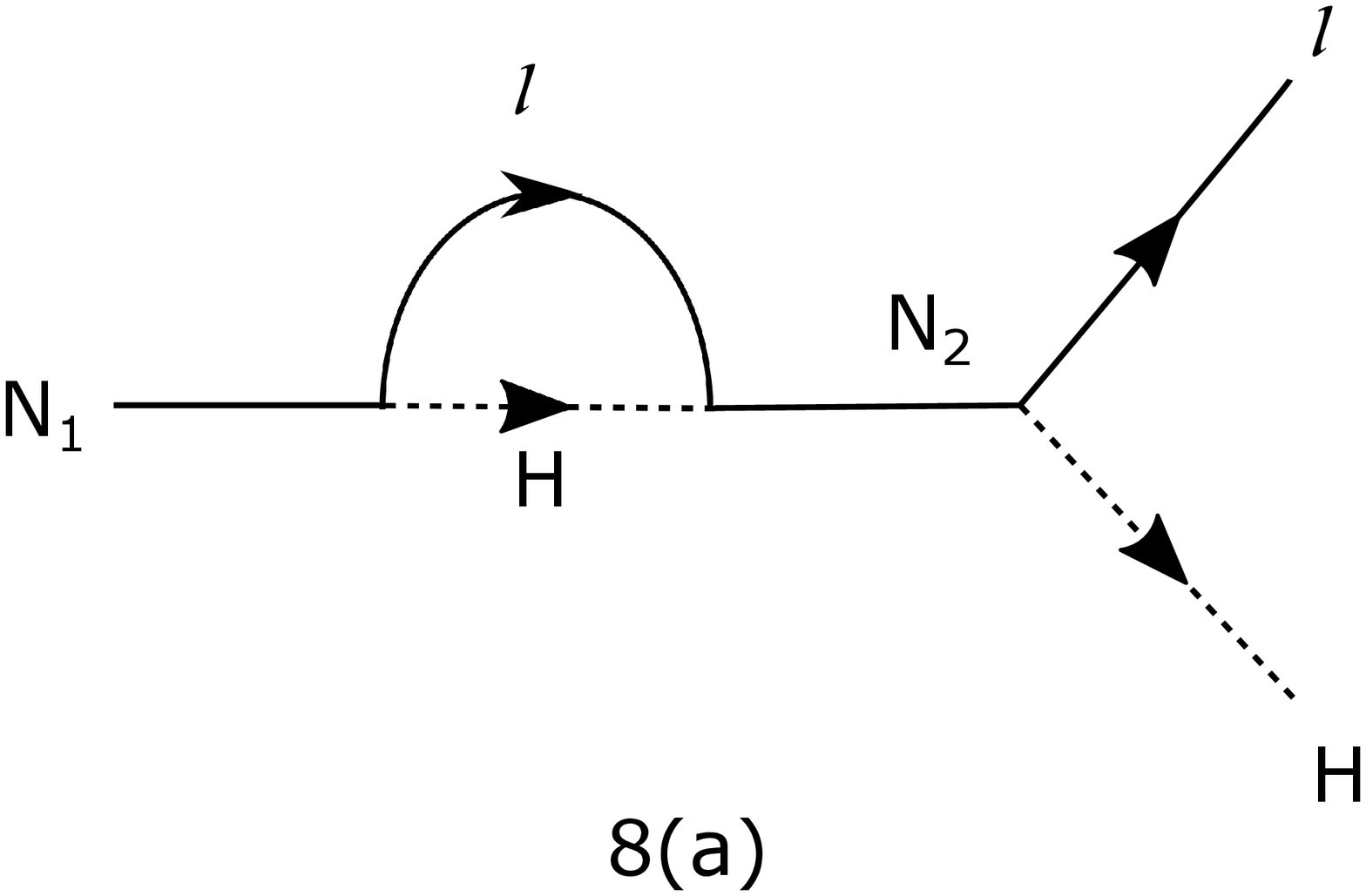}
		\includegraphics[scale=0.23]{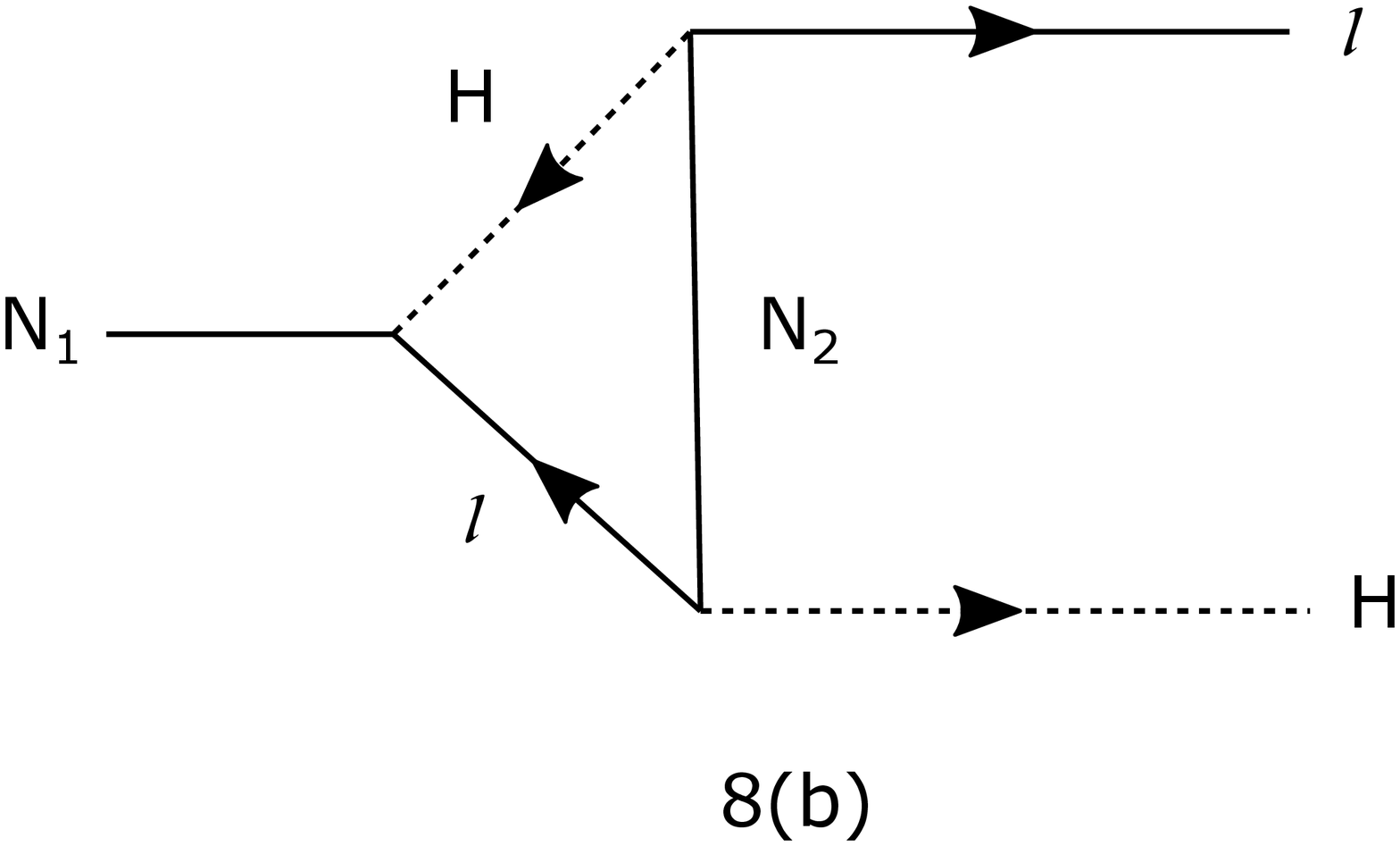}
		\includegraphics[scale=0.23]{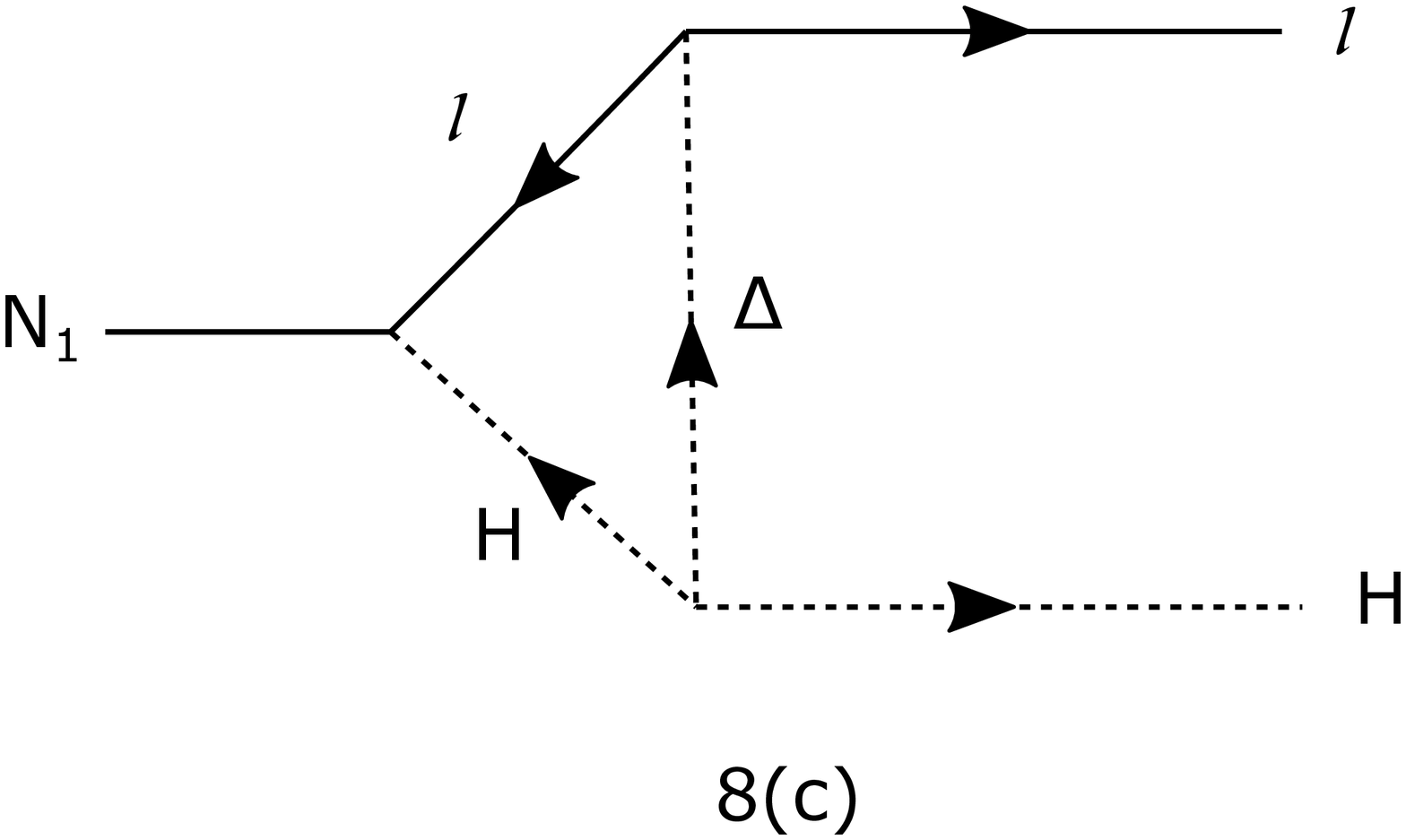}
		\caption{~\label{fig1} One loop-level diagrams contributing to $CP$ asymmetry.}
	\end{figure}

\begin{figure}[t]
    \centering
    \includegraphics[scale=0.5]{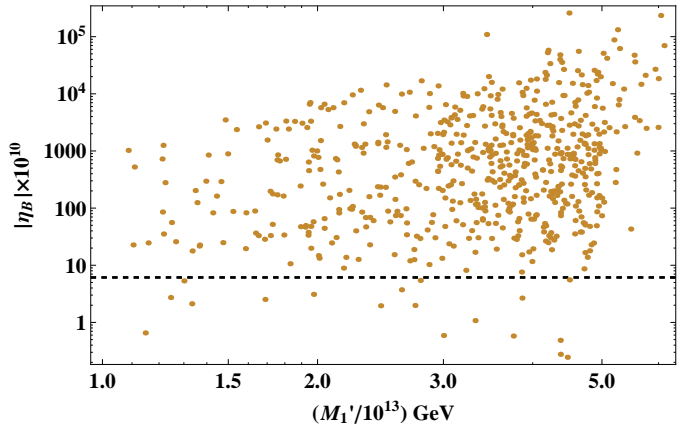}
    \caption{Variation of baryon asymmetry with lightest right-handed neutrino mass $M_{1}'$ for normal hierarchy. Dashed line denotes the value  observed baryon asymmetry $|\eta_B|=(6.12 \pm 0.04) \times 10^{-10}$.}
    \label{fig9}
\end{figure}
\begin{figure}[h]
 \begin{center}
 \epsfig{file=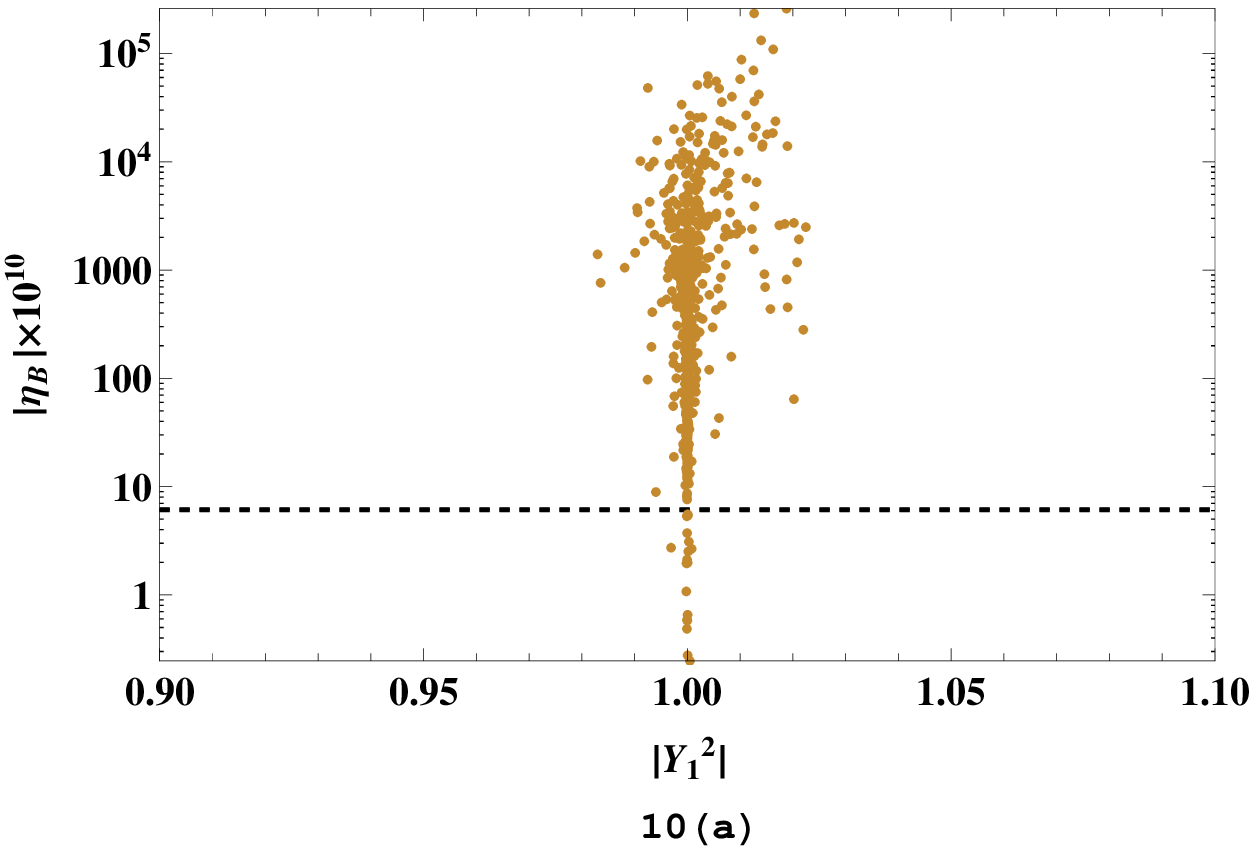,height=5.0cm,width=8.0cm}\\
 \vspace{0.5 cm}
{\epsfig{file=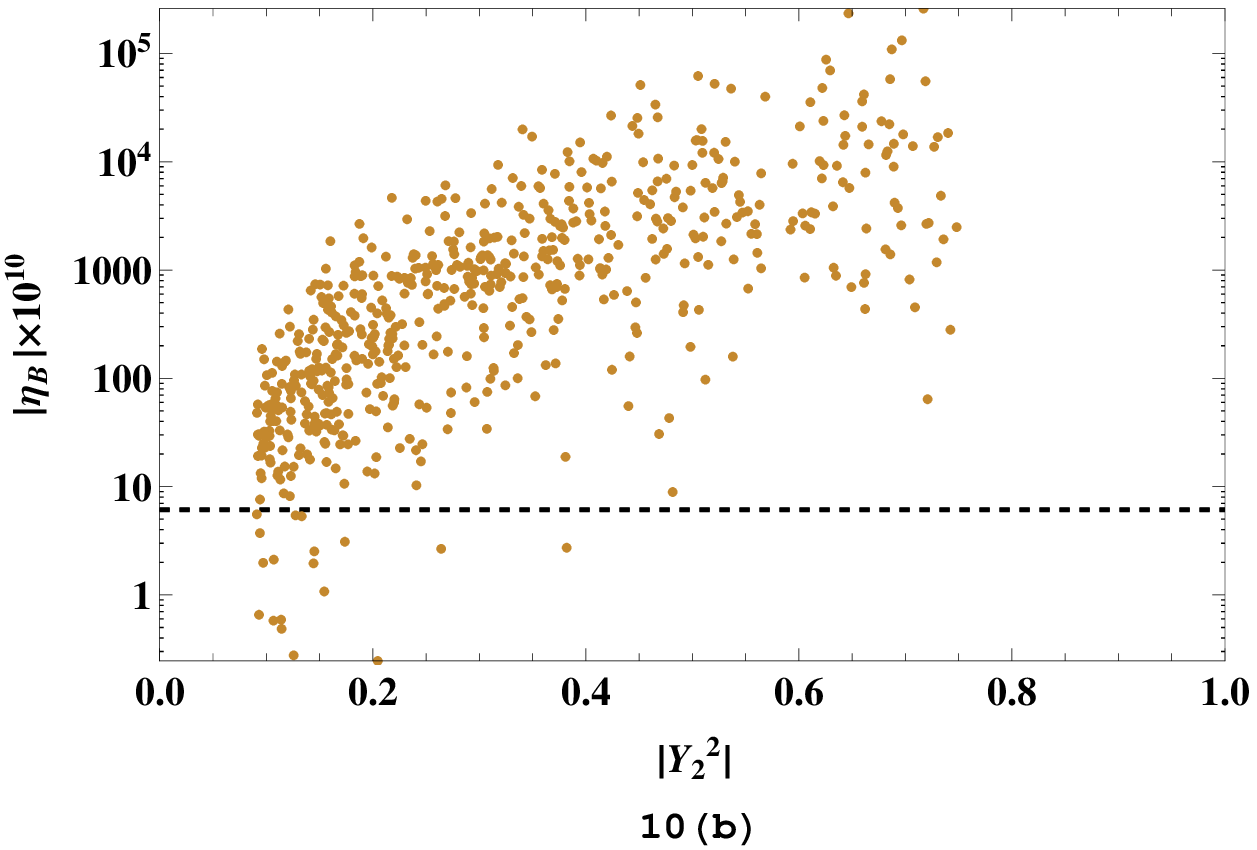,height=5.0cm,width=7.0cm},
\epsfig{file=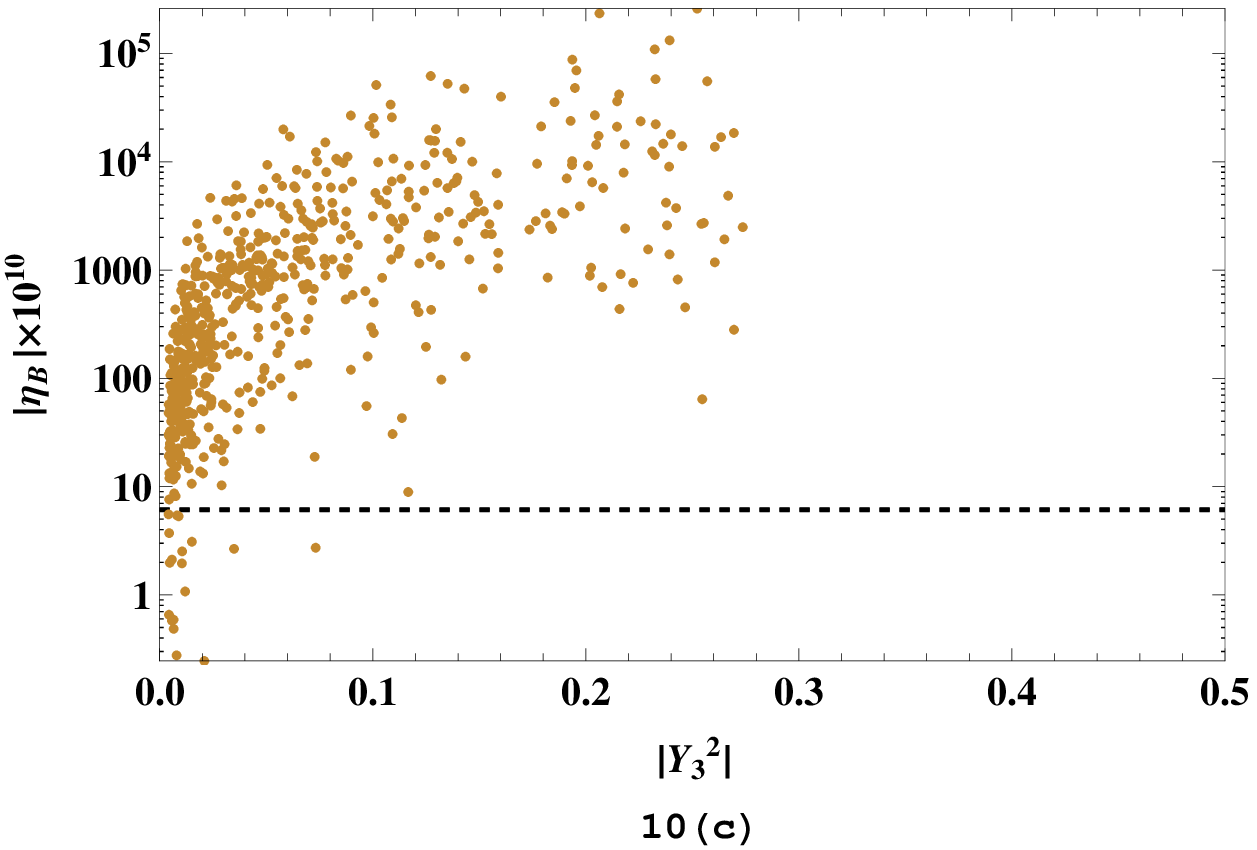,height=5.0cm,width=7.0cm}}
\end{center}
  \caption{\label{fig10}Variation of baryon asymmetry with absolute value of Yukawa couplings of modular weight 2 ($|Y_{1}^{2}|$, $|Y_{2}^{2}|$, $|Y_{3}^{2}|$) for normal hierarchy. Dashed line denotes the value  observed baryon asymmetry $|\eta_B|=(6.12 \pm 0.04) \times 10^{-10}$. } 
\end{figure}

	
Employing constraints $M_{1}^{'}<<M_{2}^{'}$ and $m^{*}>\bar{m}$ in the numerical analysis, with efficiency factor given in Eqn. (\ref{efficiency}), we estimate the baryon asymmetry of the Universe(BAU) using Eqn.(\ref{eta}). The variation of baryon asymmetry with right-handed neutrino mass is shown in Fig. (\ref{fig9}). It is evident from Fig. (\ref{fig9})  that the observed value of baryon asymmetry is consistent for right-handed neutrino mass scale in the range ($(1-5)\times10^{13}$) GeV. Also, complex modulus $\tau$ is the only source of $CP$ violation and hence is responsible for generating baryon asymmetry. It is to be noted that in modular symmetry the Yukawa couplings are not free and are, in general, holomorphic functions of modulus $\tau$, so it is imperative to investigate implication of observed BAU on $|Y_{1}^{2}|$, $|Y_{2}^{2}|$ and $|Y_{3}^{2}|$. In Fig.(\ref{fig10}), we have depicted the variation of baryon asymmetry with Yukawa couplings of modular weight 2. The Yukawa couplings $|Y_{1}^{2}|$, $|Y_{2}^{2}|$ and $|Y_{3}^{2}|$ are, further, constrained in the range ($0.98-1.02$), ($0.10-0.75$) and ($0.01-0.28$), respectively.

\section{Conclusions}
\label{sec:6}
 In conclusion, we have proposed  a modular A$_4$ symmetric model in the supersymmetric framework wherein, along with Type-I seesaw, the effective Majorana neutrino mass matrix can be generated via two scenarios \textit{viz.,} (i) allowing dimension-5 Weinberg operator and (ii) retaining the overall superpotential of the model to tree level dimension-4 where scale-breaking is accomplished via introducing scalar triplet Higgs superfields($\Delta,\bar{\Delta}$). In minimal setup of beyond standard model field content, Type-I seesaw results in scaling neutrino mass matrix giving vanishing lowest neutrino mass eigenvalue and $\theta_{13}=0$. Interestingly, the breaking patterns in both, otherwise dynamically different scenarios, are similar which can be attributed to the same charge assignments of superfields($\Delta,\bar{\Delta}$) and the Higgs superfield $H_u$ under modular $A_4$ symmetry. The breaking is found to be proportional to the Yukawa coupling of modular weight 10($Y_{1,1'}^{10}$) ameliorating the correct low energy phenomenology. Furthermore, scenario-2 has been numerically investigated to predict the correct low energy phenomenology and matter-antimatter asymmetry. We find that the model satisfies neutrino oscillation data for normal hierarchical neutrino masses. Although the model contains Yukawa couplings of different weights(2,4,6 and 8) but scale-breaking depends on Yukawa coupling of modular weight 10($Y_{1,1'}^{10}$) only. The constraint to have correct low energy phenomenology restricts Yukawa couplings of different weights to narrow ranges as shown in Figs.(\ref{fig:1}-\ref{fig:3}) which are consistent with perturbative limit $max\left[Y_m^n\right]\leq\sqrt{4\pi}$\cite{Nomura:2019lnr}. The Yukawa couplings of lowest weight (weight 2) especially $|Y_2^{2}|$ and $|Y_3^{2}|$ have sharp correlation with imaginary part of complex modulus $\tau$(Fig.(\ref{fig:1})) whereas $|Y_1^{2}|$ is insensitive to $\tau$ and thus to $CP$ violation. One of the interesting feature of the model is the appearance of lower bound on sum of neutrino masses. In fact the model predicts a very robust range for $\sum m_{i}$ ($0.05-0.08$)eV(Figs.(\ref{fig:4}) and (\ref{fig:5})). The effective Majorana mass  $M_{ee}$ is found to be in the range ($0.04-0.06$)eV. In the near future, $0\nu\beta\beta$ decay  experiments will, hopefully, reach the sensitivity of a few meV to $M_{ee}$. These experiments especially NEXT and nEXO shall provide crucial test for viability the model. Furthermore, there is, however, no decisive prediction about the possible $CP$ violation. The model is consistent with both $CP$ conserving and $CP$ violating solutions. Also, we have studied the leptogenesis in the framework of Type-I+II seesaw. For successful baryogenesis, the right-handed Majorana neutrino mass scale is to be quite high ($(1-5)\times10^{13})$GeV) which corresponds to the region where flavour effects are negligible\cite{Dev:2017trv}(Fig.(\ref{fig9})). The observed baryon asymmetry, further, constrain Yukawa couplings of modular weight 2 as shown in Fig.(\ref{fig10}). The constrained ranges of Yukawa couplings of modular weight 2 are $|Y_{1}^{2}|\rightarrow ($0.98-1.02$)$, $|Y_{2}^{2}|\rightarrow ($0.10-0.75$)$ and $|Y_{3}^{2}|\rightarrow ($0.01-0.28$)$. For ready reference, it is useful to give benchmark points satisfying the neutrino phenomenology and BAU \textit{viz}. $g_{i}(i=1,2,3,4)=(0.01,0.50,0.25,0.30),  K_{i}(i=1,2,3,4)=(0.70,0.50,0.50,0.50)$ for the complex modulus $\tau=(0.25+i 1.85)$ in the fundamental range.
\section*{Acknowledgments}
 M. K. acknowledges the financial support provided by Department of Science and Technology(DST), Science and Engineering Research Board(SERB), Government of India vide Grant No. DST/INSPIRE Fellowship/2018/IF180327. The authors, also, acknowledge Department of Physics and Astronomical Science for providing necessary facility to carry out this work.

\end{document}